\documentclass[12pt]{article}
\usepackage{amsfonts}
\usepackage{amsmath, amssymb}
\usepackage{hyperref}
\usepackage[dvips]{color}
\usepackage{graphicx, subfigure}
\usepackage{psfrag}
\usepackage{mathrsfs}

\usepackage{wrapfig}

\newcommand{\Ccal}{{\mathcal C}}

\newcommand{\Hcal}{{\mathcal H}}
\newcommand{\Ical}{{\mathcal I}}

\newcommand{\Ncal}{{\mathcal N}}

\newcommand{\Scal}{{\mathcal S}}
\newcommand{\Tcal}{{\mathcal T}}

\newcommand{\nn}{\nonumber}
\newcommand{\f}{\frac}

\def\th{{\theta}}

\def\rem#1{}
\def\al{{\alpha}}

\def\rem#1{}

\newcommand\non{\nonumber \\}
\newcommand{\bel}{\begin{eqnarray}}
\newcommand{\ee}{\end{eqnarray}}

\newcommand{\Ge}{\Gamma_e}

\newcommand{\ii}{i}

\newcommand{\vv}{v}
\newcommand{\uu}{u}
\newcommand{\zz}{z}

\newcommand{\an}{a}
\newcommand{\bn}{b}

\newcommand{\Tpp}{\mathcal{T}^{(++)}}
\newcommand{\Tbm}{\mathcal{T}^{(++)}_{\beta,-}}
\newcommand{\Tmp}{\mathcal{T}^{(-+)}}
\newcommand{\Tmpbm}{\mathcal{T}^{(-+)}_{\beta,-}}

\newcommand{\Be}{\beta}
\newcommand{\Al}{\alpha}
\newcommand{\De}{\delta}
\newcommand{\Ga}{\gamma}

\newcommand{\p}{{\bf p}}
\newcommand{\q}{{\bf q}}
\newcommand{\T}{{\bf t}}

\newcommand{\rr}{r}

\newcommand{\lr}{l}

\newcommand{\lat}{\lambda} 
\newcommand{\dd}{D}
\newcommand{\wt}{w} 
\newcommand{\vt}{v} 
\newcommand{\ut}{u}
\newcommand{\Pt}{P}
\newcommand{\Dm}{\Delta_{{\q},{\T}}}

\usepackage{color}

\renewcommand{\title}[1]{\vbox{\center\LARGE{#1}}\vspace{5mm}}
\renewcommand{\author}[1]{\vbox{\center\large#1}\vspace{5mm}}

\numberwithin{equation}{section}

\begin{document}

\begin{titlepage}

\bigskip
\hfill\vbox{\baselineskip12pt
\hbox{KIAS-P16040}
}
\bigskip\bigskip\bigskip\bigskip\bigskip\bigskip\bigskip\bigskip

\begin{center}
\Large{ \bf
Superconformal index with surface defects for class ${\cal S}_k$ 
}
\end{center}
\bigskip
\bigskip

\bigskip
\bigskip
\centerline{ \large  Yuto Ito and Yutaka Yoshida}
\bigskip
\bigskip
\bigskip
\centerline{  \it School of Physics, Korea Institute for Advanced Study (KIAS),}
\centerline{\it  85 Hoegiro Dongdaemun-gu, Seoul, 02455, Republic of Korea. }
\centerline{yutoito1986316ATgmail.com, \ \  yyyyosidaATgmail.com}
\bigskip
\bigskip

\begin{abstract}
We study surface defects in 4d $\mathcal{N}=1$  $SU(N)$ superconformal gauge theories of class $\mathcal{S}_k$ obtained from the 6d (1,0) theories of type $A_{N-1}$, which are worldvolume theories on $N$ M5-branes at $\mathbb{C}^2/\mathbb{Z}_k$ singularities, compactified on Riemann surfaces with punctures.
 First we apply a method based on Riemann surface description and obtain the superconformal index of the theories in the presence of surface defects  labelled by arbitrary symmetric representations of $su(N)$.
Then we propose another description for the same surface defects, which involves 4d-2d coupled systems, by identifying which 2d $\mathcal{N}=(0,2)$ theories should be coupled. We compute the index of the 4d-2d systems and reproduce the results obtained from the first method.
Finally we study the 2d TQFT structure of the index for class $\mathcal{S}_{k}$ theories by obtaining several eigenfunctions and eigenvalues of the difference operators that capture the surface defects and checking their relation.
\end{abstract}
\end{titlepage}

\newpage
\baselineskip=18pt

\tableofcontents

\section{Introduction}
There are interesting classes of 4d superconformal field theories (SCFTs), which can be obtained by compactifying 6d SCFTs on Riemann surfaces with punctures.
The first example is a class of 4d $\Ncal=2$ SCFTs that descend from the 6d (2,0) SCFTs of type $A_{N-1}$, which are worldvolume theories of $N$ M5-branes. This class is called class $\Scal$ and theories of the class were classified in terms of Riemann surfaces in \cite{Gai1}.
Since the choice of punctured Rieman surfaces determines the obtained 4d SCFTs, we can label each of these theories by a certain Rieman surface.
Furthermore, gluing two Riemann surfaces amounts to coupling the two associated SCFTs by using $\Ncal=2$ vector multiplets.

Inspired by the relation between the 4d SCFTs and the Riemann surfaces, the authors of \cite{GPRR} and \cite{GRRY} introduced a systematic formalism that computes superconformal index of these 4d SCFTs in terms of the Riemann surface description. 
This formalism allows us to obtain the index even for theories without known Lagrangians.
In this sense, it is useful to associate SCFTs to Riemann surfaces and find other classes of theories that admit a similar Riemann surface description.

Recently other classes of 4d SCFTs are associated to Riemann surfaces in \cite{Gaiotto:2015usa}. They are 4d $\Ncal=1$ theories that are obtained by the compactification of the 6d (1,0) SCFTs of type $A_{N-1}$, the worldvolume theories on $N$ M5-branes at the singularities $\mathbb{C}^2/\mathbb{Z}_k$, on the Riemann surfaces. These classes are called class $\Scal_k$. Their generalization was studied in \cite{FHU, HM}. 
As in the case of class $\Scal$ theories,
the authors of \cite{Gaiotto:2015usa} developed a similar formalism that computes superconformal index of the class $\Scal_k$ theories from the Riemann surfaces.

As a further investigation of these theories,  we study half-BPS surface defects \cite{GW,GW2,Gaiotto:2009fs} in the class $\Scal_k$ theories in this paper. The Riemann surface description is again useful to compute the index in the presence of surface defects. 

First let us recall how to describe half-BPS surface defects in class $\Scal$ theories. Here we restrict to the surface defects that descend from codimension-four defects in 6d (2,0) $A_{N-1}$ SCFTs. They are labelled by irreducible representations of ${\it su(N)}$. In this paper we only consider the symmetric representations, which are associated to positive integers $r$.

There are two known methods to calculate the index of class $\Scal$ theories with the surface defects. The first one was introduced in \cite{Gaiotto:2012xa} inspired by the Riemann surface description.  
From the viewpoint of M-theory, $N$ M5-branes wrapping a Riemann surface give rise to a 4d $\Ncal=2$ $SU(N)$ SCFT and adding M2-branes at a point on the Riemann surface to this system engineers the surface defects in the 4d theory. 
Therefore, to describe the surface defects, the authors of \cite{Gaiotto:2012xa} inserted an additional puncture in the Riemann surface description and it corresponds to coupling an additional hypermultiplet to the theory in question by using $\Ncal=2$ vector multiplets.
 Let us refer to the resulting theory as a larger theory.
If we turn on a constant vacuum expectation value (vev) for the scalar field in the added hypermultiplet and look at the physics below the energy scale set by the vev, we flow to the original theory in the IR. Instead if we turn on a position-dependent vev that corresponds to a vortex configuration in the UV, we obtain the original theory in the presence of a surface defect in the IR. The field configuration of vortex number $r$ gives rise to the surface defect labelled by the $r$-th symmetric representation. 
This construction of the surface defects allows us to compute the index of class $\Scal$ theories with the surface defects from that of the larger theories without surface defects.

The second method to compute the surface defects was provided in \cite{Gadde:2013ftv}. It is based on Type IIA realization of the 4d theories. If we restrict to $SU(N)$ linear quiver theories of class $\Scal$, they can be realized by NS5-branes and D4-branes systems. Inserting the surface defects labelled by the integer $r$ corresponds to adding $r$ number of D2-branes to these systems. Noting that the worldvolume theories on the D2-branes are 2d $\Ncal=(2,2)$ $U(r)$ gauge theories, the authors of \cite{Gadde:2013ftv} first computed the index, a.k.a. the elliptic genus of the 2d theories. Then they inserted it into the contour integral expression for the index of the 4d theories in question and obtained the index with the surface defects. Moreover they checked that their result matches with the one obtained from the first method above, up to an overall fractional shift of fugacities of an $SU(N)$ flavor symmetry.

Now let us turn to class $\Scal_k$ theories. 
The authors of \cite{Gaiotto:2015usa} developed a formalism to compute the index of class $\Scal_k$ theories in the presence of the surface defects, which is generalization of the first method for class $\Scal$ theories. 
Each $\Ncal=2$ multiplet in the method for the latter theories is now reduced to $\Ncal=1$ multiplets by projecting out components that are not invariant under the $\mathbb{Z}_k$ orbifold action. For instance, when one couples extra multiplets to engineer surface defects in class $\Scal_k$ theories, one uses the orbifolded $\Ncal=2$ vector multiplets, which consist of $\Ncal=1$ vector multiplets and certain remnants of the adjoint chiral multiplets in the original $\Ncal=2$ vector multiplet.
From this formalism, they obtained the result for the surface defect labelled by the integer $r=1$ in $SU(2)$ gauge theories of class $\Scal_{k=2}$. In this paper we extend their result to the surface defects labelled by generic positive integers $r$ in $SU(N)$ gauge theories.

Alternatively, as discussed in \cite{Gaiotto:2015usa}, we can use pure $\Ncal=1$ vector multiplets instead of the orbifolded $\Ncal=2$ vector multiplets \cite{FHU,HM} when we couple the extra multiplets. It gives rise to a different type of surface defects, which are again labelled by positive integers $r$. The difference between the two types can be encoded as $\pm$ signs of extra Riemann surfaces that we glue to insert the surface defects.\footnote{These signs of building blocks were first introduced for theories obtained by compactification of the 6d (2,0) theories in \cite{BG} and also considered for 4d theories coming from the 6d (1,0) theories in \cite{HM}.} In this paper, we also calculate the index with this second type of surface defects explicitly by gluing extra Riemann surfaces and obtain the results for generic positive integers $r$.

On the other hand, there should be another method to describe the above surface defects in terms of 4d-2d coupled systems, as in the case of class $\Scal$ theories. In this paper we propose which 2d $\mathcal{N}=(0,2)$ theories should be coupled to the 4d theories in order to describe the surface defects of the above two types
 in 4d $SU(N)$ gauge theories of class $\Scal_{k}$. Moreover we calculate the index of the 4d-2d coupled systems and check that the resulting index reproduces the one computed from the formalism in \cite{Gaiotto:2015usa}, up to an overall fractional shift of fugacities of $SU(N)$ flavor symmetries as in the class $\Scal$ case. 
 
We also discuss the 2d topological quantum field theory (TQFT) structure of the index for class $\Scal_k$ theories.
It was found that the index of class $\Scal$ theories can be written in terms of correlation functions of a certain TQFT on the corresponding Riemann surface $\Ccal$ in \cite{GRRY,GRRY2}. 
It can be understood as follows.
Let us recall that class $\Scal$ theories are obtained by the compactification of the twisted 6d (2,0) theory on $\Ccal$. On the other hand, we can compactify this 6d theory on $S^1\times S^3$ and obtain a 2d theory on $\Ccal$. 
If we focus on quantities in the 4d theories and the 2d theory that come from the same protected observables in the 6d theory, there should be relation between these quantities. 
Since the 4d index does not depend on the coupling constant, the corresponding quantity should be independent of the complex structure of the Riemann surface $\Ccal$. Hence the corresponding 2d theory is a TQFT. Moreover the structure constants in the TQFT representation of the index are diagonalized by the eigenfunctions of the difference operators that capture the surface defects in class $\Scal$ theories \cite{Gaiotto:2012xa}.

Similarly we assume that the index of class $\Scal_k$ theories also has the 2d TQFT structure with diagonal structure constants. 
In this paper, we 
obtain several eigenfunctions and their eigenvalues of the difference operators for the surface defects in class $\Scal_{k=2}$ theories as a continuation of the study in \cite{Gaiotto:2015usa}. 
We check that they satisfy a relation coming from the assumption of the 2d TQFT structure.

The organization of this peper is as follows. In Section \ref{sec:Sk}, we review 4d $\Ncal=1$ linear quiver theories of class $\Scal_k$ and how to associate them to Riemann surfaces. In Section \ref{sec:Riem}, we calculate the index of these theories in the presence of the surface defects by means of the formalism 
based on the Riemann surface description and present the results for the surface defects labelled by generic positive integers. 
In Section \ref{sec:2d4d}, 
we identify the 2d theories which should be coupled to 4d theories of class $\Scal_k$ in order to capture the surface defects in these 4d theories. Also we check the resulting index matches with the one obtained in Section \ref{sec:Riem}.
In Section \ref{sec:tqft}, we obtain the first several eigenfunctions and their eigenvalues of the difference operators describing the surface defects and find that they satisfy the relation that can be derived from the assumption of the 2d TQFT structure.

While we are finishing this paper, a related work \cite{MY} has appeared.

\section{Class $\Scal_k$ theories}
\label{sec:Sk}

\begin{figure}[tb]
\begin{center}
\includegraphics[width=16cm]{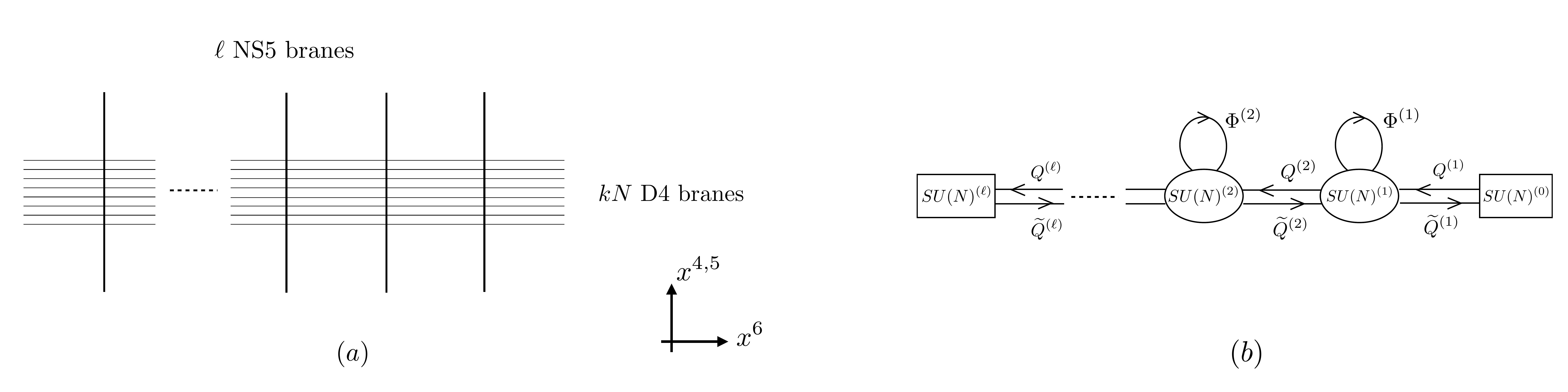}
\end{center}
\vspace{-0.5cm}
\caption{$(a)$: Brane configuration for 4d $\Ncal=2$ linear quiver gauge theories. $(b)$: The quiver diagram for the theories from the brane configuration. It is drawn in terms of 4d $\Ncal=1$ chiral multiplets $\Phi^{(m)}, Q^{(m)}$ and $\widetilde{Q}^{(m)}$ for $m=1,\cdots,\ell-1$ or $\ell$.}
\label{fig:T1kNl}
\end{figure}

First we review class $\Scal_k$ theories, which are certain 4d $\Ncal =1$ SCFTs and were associated to Riemann surfaces in \cite{Gaiotto:2015usa}. In this paper we restrict to $SU(N)$ linear quiver gauge theories of this class. These theories are obtained by an orbifold projection on the worldvolume theory on the Type I\vspace{-0.2em}IA brane configuration in Figure \ref{fig:T1kNl}$(a)$.  The $\ell$ NS5-branes and the $kN$ D4-branes extend along the directions $012345$  and the directions $01236$ respectively.

Before orbifolding, the worldvolume theory on the D4-branes are 4d $\Ncal=2$ gauge theories. We denote this theory by $\Tcal_{1,kN,\ell}$ and 
describe its quiver diagram in Figure \ref{fig:T1kNl}$(b)$ in terms of $\Ncal=1$ language.
The circular nodes contain $\Ncal=1$ vector multiplets and $\Phi^{(m)}$ where $m=1,\cdots, \ell-1$ denote massless adjoint chiral multiplets.  
In addition, the horizontal lines correspond to the  chiral multiplets $Q^{(m)}$ and $\widetilde{Q}^{(m)}$ $(m=1,\cdots,\ell)$.
The chiral multiplets $Q^{(m)}$ transform as fundamental and anti-fundamental under $SU(kN)^{(m)}$ and $SU(kN)^{(m-1)}$ respectively. Similarly $\widetilde{Q}^{(m)}$ do as the anti-fundamental and fundamental representation under these groups respectively. They couple through the cubic superpotential $W=\sum_m (Q^{(m)} \Phi^{m} \widetilde{Q}^{(m)}-\widetilde{Q}^{(m+1)} \Phi^{(m)} Q^{(m+1)})$, where a trace over gauge indices is implicitly taken.

Now we consider the orbifold $\mathbb{C}^2/\mathbb{Z}_k$.
It acts on the $45$ and $89$ planes as rotation in opposite directions 
\begin{align}
x^4+ix^5 \rightarrow e^{\f{2\pi i}{k}} (x^4+ix^5) \, , \ \ \ 
x^8+ix^9 \rightarrow e^{-\f{2\pi i}{k}} (x^8+ix^9) \, .
\label{eq:orbifold}
\end{align}
This is an element of $\mathbb{Z}_k$ discrete subgroup of the rotation generated by $h_{45}-h_{89}$ and we denote it as $\mathbb{Z}_k^{({\rm R})}$. Recalling that the $U(1)_r$ and $SU(2)_R$ R-charges correspond to $-h_{45}$ and $h_{89}$ respectively, we see that $\Phi^{(m)}$ get $\mathbb{Z}_k$ charge $1$ and $Q^{(m)}$ and $\widetilde{Q}^{(m)}$ get $\mathbb{Z}_k$ charge $-\f{1}{2}$ from $\mathbb{Z}_k^{({\rm R})}$ action.

In addition, each of the $SU(kN)^{(m)}$ groups has $\mathbb{Z}_k$ discrete subgroup whose generator acts on the fundamental representation as diag$(\alpha^{\f{m}{2}}1_N,\alpha^{\f{m}{2}+1} 1_N, \alpha^{\f{m}{2}+2} 1_N, \cdots)$ with $\alpha=e^{2\pi i/k}$. We denote it as $\mathbb{Z}_k^{(m)}$ and identify the diagonal product of these $\mathbb{Z}_k^{({\rm R})}$ and $\mathbb{Z}_k^{(m)}$ actions with the orbifold action. After orbifolding, we only have the multiplets invariant under it.

\begin{figure}[tb]
\begin{center}
\includegraphics[width=17cm]{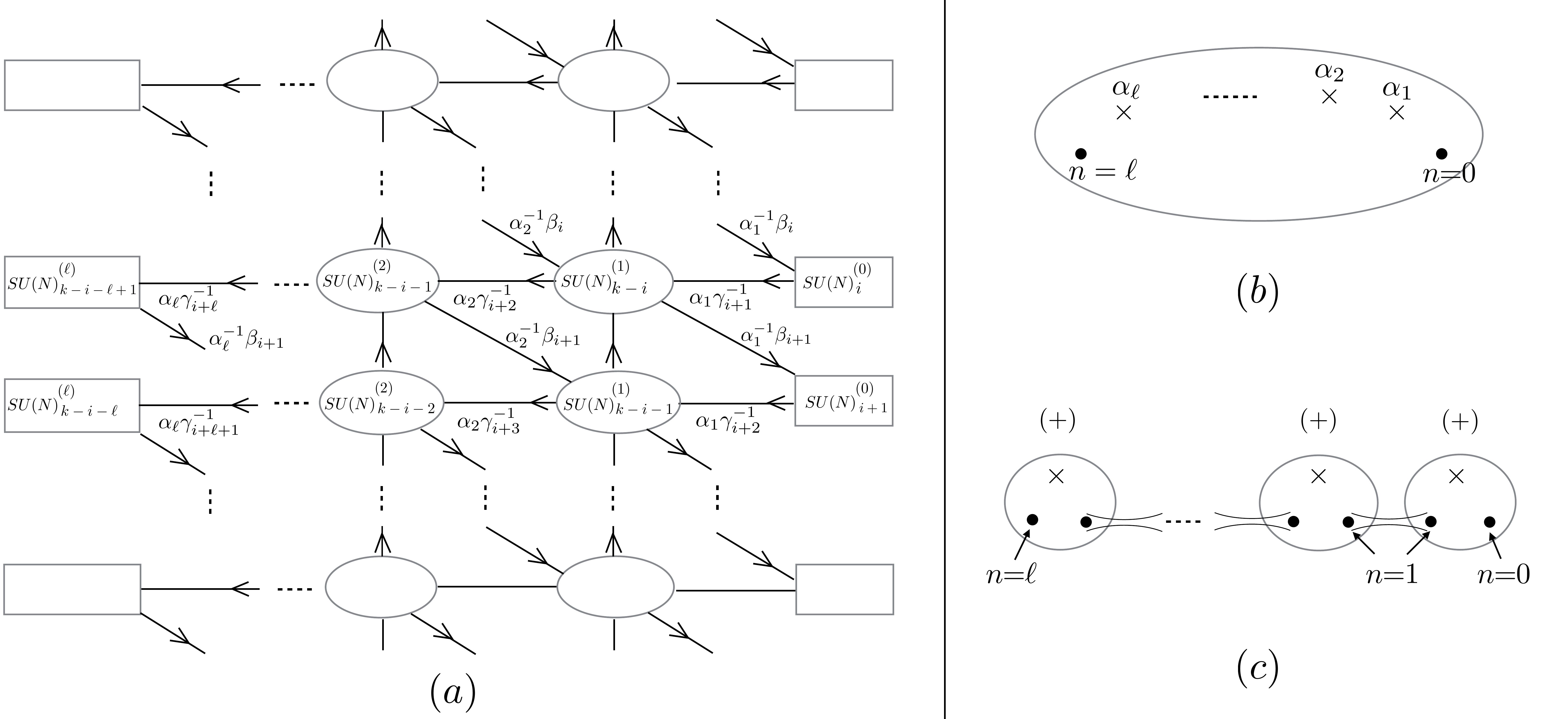}
\end{center}
\vspace{-0.5cm}
\caption{$(a)$: The quiver diagram for the theory $\Tcal_{k,N,\ell}$. Each arrow corresponds to an $\Ncal=1$ chiral multiplet. 
The parameters $\alpha_m, \Be_{\ii},\Ga_{\ii}$ are the fugacities for the $U(1)$ flavor symmetries and their powers indicate the flavor charges of each chiral multiplet.
$(b)$: The Riemann surface 
on which we compactify the 6d theory to obtain $\Tcal_{k,N,\ell}$. The two dots represent maximal punctures of the color $n=0$ and $n=\ell$ respectively. The x's marks denote minimal punctures, each of which corresponds to each of the flavor symmetries $U(1)_{\alpha_m}$.
$(c)$: The Riemann surface in $(b)$ is obtained by gluing $\ell$ copies of a Riemann surface with two maximal punctures and one minimal puncture. Maximal punctures of the same color $n$ can be glued. Each Riemann surface has a $+$ sign.  
}
\label{fig:TkNl}
\end{figure}

The resulting 4d gauge theories on the D4-branes are described in Figure \ref{fig:TkNl}$(a)$. We denote these theories by $\Tcal_{k,N,\ell}$. 
As shown in the figure, the $SU(kN)^{(m)}$ vector multiplets in $\Tcal_{1,kN,\ell}$ are decomposed into $k$ vector multiplets with $SU(N)_{\ii}^{(m)}$, where $\ii$ runs from $1$ to $k$ modulo $k$. The $\ii$-th diagonal component of diag$(\alpha^{\f{m}{2}}1_N,\alpha^{\f{m}{2}+1} 1_N, \alpha^{\f{m}{2}+2} 1_N, \cdots)$ corresponds to the $SU(N)^{(0)}_{\ii}$ group node for $m=0$ and the $SU(N)^{(m)}_{k-\ii+m+1}$ group node for $m\geq 1$.
 The vertical lines connecting from the $SU(N)^{(m)}_{k-\ii-m}$ gauge node to the $SU(N)^{(m)}_{k-\ii-m+1}$ node represent the chiral multiplets $\Phi^{(m,\ii)}$.
 They get $\mathbb{Z}_k$ charges $\f{m}{2}+\ii-1$ and $-(\f{m}{2}+\ii)$ from $\mathbb{Z}_k^{(m)}$ action.
Summing over the $\mathbb{Z}_k$ charges from the $\mathbb{Z}_k^{(m)}$ actions and the $\mathbb{Z}_k^{({\rm R})}$ action, we see that all the chiral multiplets $\Phi^{(m,\ii)}$ drawn in the figure are invariant under the orbifold action.
The horizontal line coming to the $SU(N)^{(m)}_{k-\ii-m+1}$ group node represents the chiral multiplet $Q^{(m,\ii)}$ and the tilted line going out from the same node is $\widetilde{Q}^{(m,\ii)}$. They are also invariant under the orbifold action.

In addition to the flavor symmetries $SU(N)^{(0)}_{\ii}$ and $SU(N)^{(\ell)}_{\ii}$, these theories have flavor symmetries $\prod_{m=1}^{\ell}U(1)_{\alpha_m} \times U(1)_{\T} \times \prod_{\ii=1}^{k} U(1)_{\beta_{\ii}} \times  \prod_{\ii=1}^{k} U(1)_{\gamma_{\ii}}$ with the constraint that the sum of all $U(1)_{\beta_{\ii}}$ charges is zero and the same for $U(1)_{\gamma_{\ii}}$.

The charges of the chiral multiplets 
$Q^{(m,\ii)}$ and $\widetilde{Q}^{(m,\ii)}$
under the flavor symmetries $U(1)_{\Al_m},$ $U(1)_{\Be_{\ii}},$ $U(1)_{\Ga_{\ii}}$ are represented in Figure \ref{fig:TkNl}$(a)$. The powers of the parameters $\Al_m, \Be_i$ and $\Ga_i$ indicate these charges. In addition, all of these multiplets have $U(1)_{\T}$ charge $\f{1}{2}$. 
After orbifolding, the cubic superpotential above is reduced to
\begin{align}
&W= \sum_{m,\ii} Q^{(m,\ii+1)} \Phi^{(m,\ii)} \widetilde{Q}^{(m,\ii)}- \widetilde{Q}^{(m+1,\ii)} \Phi^{(m,\ii)} Q^{(m+1,\ii)} \, .
\end{align}
From this superpotential, we can read off the charge of $\Phi^{(m,\ii)}$ as ${\T}^{-1} \Be^{-1}_{\ii+1} \Ga_{\ii+m+1}$.

Let us turn to the M-theory perspective.
The theories $\Tcal_{1,N,\ell}$ have interpretation as the compactification of the 6d (2,0) SCFTs of type $A_{N-1}$ on Riemann surfaces with punctures \cite{Gai1}.
As its generalization to the theories $\Tcal_{k,N,\ell}$,
it is natural to consider that they are obtained from the compactification of the 6d (1,0) $A_{N-1}$ SCFTs, which are worldvolume theories on M5-branes at the $\mathbb{C}^2/\mathbb{Z}_k$ singularity. In order to obtain the quiver gauge theory in Figure \ref{fig:TkNl}$(a)$, we compactify the 6d theories on the Riemann surface with two maximal punctures and $\ell$ minimal punctures described in Figure \ref{fig:TkNl}$(b)$. This Riemann surface can be constructed from gluing $\ell$ building blocks, called trinions, as described in Figure \ref{fig:TkNl}$(c)$. Each of these trinions has two maximal punctures and one minimal puncture. 
To each maximal puncture, we assign a color $n\ (n=0,1,\cdots, \ell\ {\rm modulo}\ k)$  to encode $U(1)_{\beta_{\ii}}$ charges of the chiral multiplets. The rightmost trinion, whose two maximal punctures have colors $n=0$ and $1$ respectively, corresponds to the theory $\Tcal_{k,N,1}$ and the $m$-th one from the right corresponds to $\Tcal^{(m)}_{k,N,1}$, which we can obtain by replacing $\Be_i$ and $\Al_1$ in $\Tcal_{k,N,1}$ with $\Be_{i-(m-1)}$ and $\Al_m$. Moreover a sign $+$ is associated to each trinion. As in \cite{HM}, a sign $\pm$ of a trinion indicates the directions of the NS5-brane corresponding to its minimal puncture. Later we consider trinions with $-$ signs.

The maximal punctures of the same color can be glued to each other. When we glue the two theories $\Tcal_{k,N,1}$ and $\Tcal_{k,N,1}^{(2)}$, we gauge diagonal combination of $SU(N)^{(1)}_{k-i}$ in the former and $SU(N)^{(0)}_{i+1}$ in the latter. In addition, we add bifundamental chiral multiplets $\Phi$, which connect vertically neighbouring gauge nodes in Figure \ref{fig:TkNl}$(a)$. Thus we obtain the theory $\Tcal_{k,N,2}$.

\begin{figure}[tb]
\begin{center}
\includegraphics[width=15cm]{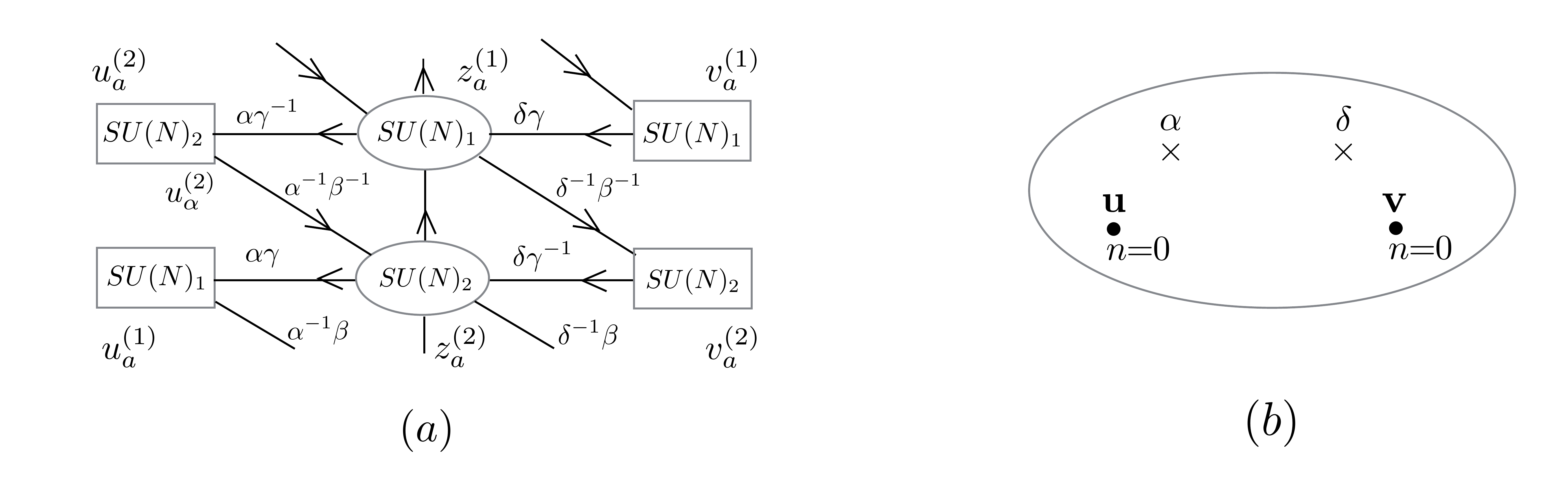}
\end{center}
\vspace{-0.5cm}
\caption{$(a)$: The quiver diagram for the theory $\Tpp$. The parameters $\uu^{(\ii)}_{\an}, \zz^{(\ii)}_{\an}, \vv^{(\ii)}_{\an}\ (\ii=1,2$ and $a=1,\cdots, N)$ are the fugacities for the Cartans of each $SU(N)_{\ii}$ gauge groups and flavor symmetry groups. $(b)$: The Riemann surface that corresponds to $\Tpp$. The parameters associated to the maximal punctures have two components, namely ${\bf \uu}=(\uu^{(1)}_{\an},\uu^{(2)}_{\an})$ and ${\bf \vv}=(\vv^{(1)}_{\an},\vv^{(2)}_{\an})$. }
\label{fig:T2N2}
\end{figure}
Next we give the superconformal index for the theory $\Tpp$ in Figure \ref{fig:T2N2} on $S^1\times S^3$. This theory can be obtained by replacing $\Al_1$ and $\Al_2$ in $\Tcal_{2,N,2}$ with $\De$ and $\Al$ respectively. 
The index is defined as a trace over the Hilbert space $\Hcal_{S^3}$ on $S^3$ \cite{Romelsberger:2005eg,Kinney:2005ej}.
\begin{align}
\Ical={\rm Tr}_{\Hcal_{S^3}} (-1)^F {\p}^{j_1+j_2-\f{r}{2}} {\q}^{j_1-j_2-\f{r}{2}} \prod_{\ell \in \mathfrak{F}} f_{\ell}^{q_{\ell}} \, ,
\end{align}
where $j_1$ and $j_2$ are the Cartans of the $SO(4)\sim SU(2)_1\times SU(2)_2$ isometry of $S^3$ and $r$ is the $U(1)_r$ R-symmetry. The charges $q_{\ell}$ are for the flavor symmetries, the set of which is denoted by $\mathfrak{F}$ and their fugacities are $\{f_{\ell}\}=\{\alpha_m, {\T}, \beta_{\ii}, \gamma_{\ii},\uu^{(\ii)}_{\an},\vv^{(\ii)}_{\bn}\}$, where $m=1,\cdots,\ell, \  \ii=1,2$ and $\an,\bn=1,\cdots,N$. The fugacities $\uu^{(\ii)}$ and $\vv^{(\ii)}$ correspond to $SU(N)_i$ flavor symmetries described as the left and right nodes in Figure \ref{fig:T2N2}$(a)$ respectively. Thus we have $\prod_{\an=1}^N \uu^{(\ii)}_{\an}=\prod_{\an=1}^N \vv^{(\ii)}_{\an}=1$. We denote $\Be_1$ and $\Ga_1$ as $\Be$ and $\Ga$ from now on. Using the elliptic Gamma function $\Ge(\zz):=\Gamma(\zz;{\p},{\q})$ defined in (\ref{defGam}), we give the index for $\Tpp$ as follows
\begin{align}
&\Ical_{\Tpp}=
\left( \f{\Ical_V^{N-1}}{N!} \right)^2
 \oint
\prod_{{\an},{\bn}=1}^{N-1}
\f{d
\zz^{(1)}_{\an}
}{2\pi i
\zz^{(1)}_{\an}
}
\f{d
\zz^{(2)}_{\bn}
}{2\pi i
\zz^{(2)}_{\bn}
}
\f{
\prod_{{\an},{\bn}=1}^N
\Ge(\f{\p\q}{\T}\Be\Ga \zz^{(1)}_\an (\zz^{(2)}_{\bn})^{-1}) 
 \Ge(\f{\p\q}{\T} (\Be\Ga)^{-1} (\zz^{(1)}_\an)^{-1} \zz^{(2)}_{\bn} )
}{
\prod_{\an\neq {\bn}} 
\Ge(\zz^{(1)}_\an/ \zz^{(1)}_{\bn})  
\Ge(\zz^{(2)}_\an/ \zz^{(2)}_{\bn})
}
\nonumber\\
&
\prod_{\an,{\bn}=1}^N 
\Ge(\T^{\f{1}{2}} \zz^{(1)}_{\an} (\vv^{(1)}_{\bn})^{-1} \De\Ga )
 \Ge(\T^{\f{1}{2}} \vv^{(2)}_{\an}  (\zz^{(1)}_{\bn})^{-1} \De^{-1}\Be^{-1} )
 \Ge(\T^{\f{1}{2}} \zz^{(2)}_{\an} (\vv^{(2)}_{\bn})^{-1} \De\Ga^{-1} )
 \Ge(\T^{\f{1}{2}}  \vv^{(1)}_{\an} (\zz^{(2)}_{\bn})^{-1} \De^{-1}\Be ) \nonumber\\
&
\Ge(\T^{\f{1}{2}}  (\uu^{(1)}_{\an})^{-1}  \zz^{(1)}_{\bn} \Al^{-1}\Be )
 \Ge(\T^{\f{1}{2}} \Ga^{-1} \Al  \uu^{(2)}_{\an}  (\zz^{(1)}_{\bn})^{-1})
 \Ge(\f{\T^{\f{1}{2}}}{\Be\Al} (\uu^{(2)}_{\an})^{-1}  \zz^{(2)}_{\bn}  )
 \Ge(\T^{\f{1}{2}} \Ga\Al \uu^{(1)}_{\an} (\zz^{(2)}_{\bn})^{-1}  ) \, ,
\label{eq:indT2N2}
\end{align} 
where $\Ical_V:=\prod_{n\geq 1} (1-{\p}^n)(1-{\q}^n)$ and $\prod_{\an=1}^N \zz^{(\ii)}_{\an}=1$. Here we take the R-charges of the bifundamental chiral multiplets to be $2$ and those of the other chiral multiplets to be $0$.\footnote{Applying the procedure called a-maximization, we obtain the R-charges of all the chiral multiplets as $\f{2}{3}$. If we shift ${\T}$ to ${\T}({\p}{\q})^{\f{2}{3}}$ at the end of our calculation, we will obtain correct results.}

\section{Surface defects from Riemann surface description}
\label{sec:Riem}

In this section we review one method to calculate the superconformal index for theories $\Tcal_{2,N,\ell}$ in the presence of helf-BPS surface defects, which was developed in \cite{Gaiotto:2015usa} as generalization of the work \cite{Gaiotto:2012xa} for class $\Scal$ theories. In this paper we restrict to the surface defects that fill the temporal $S^1$ and the maximal circle inside the $S^3$ fixed by the $j_1-j_2$ rotation.
\begin{figure}[tb]
\begin{center}
\includegraphics[width=14cm]{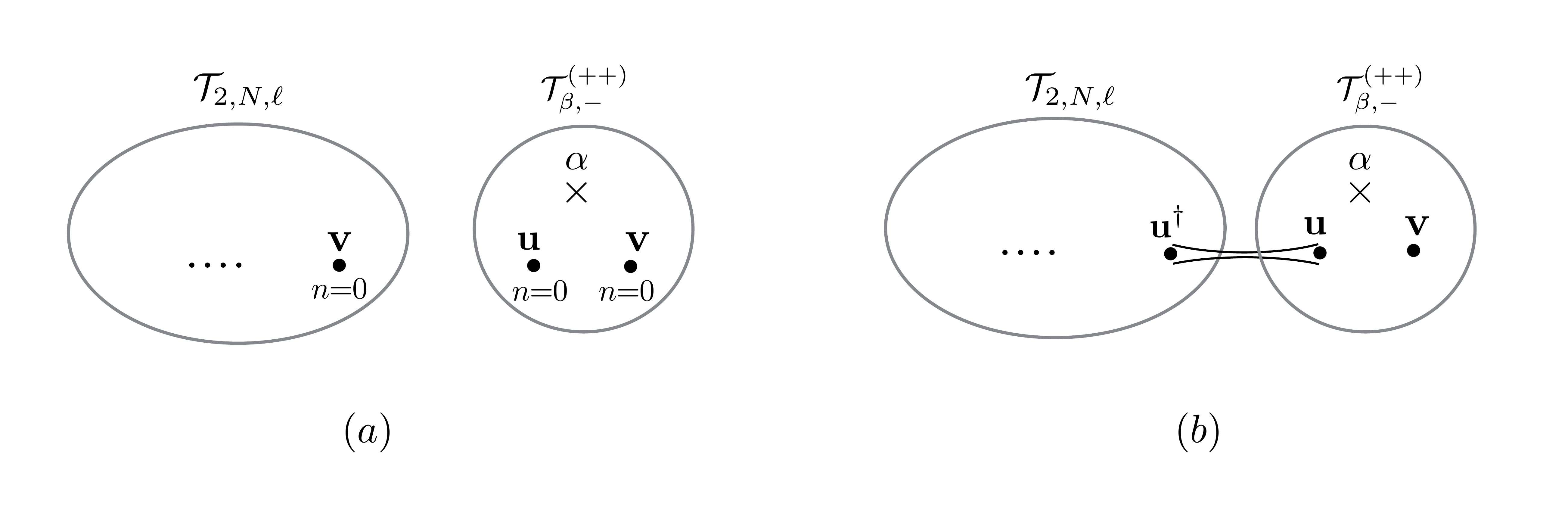}
\end{center}
\vspace{-0.5cm}
\caption{Procedure to obtain the index with a surface defect. $(a)$: One of the maximal punctures in $\Tcal_{2,N,\ell}$ has the color $n=0$ and corresponds to the $SU(N)^2$ flavor symmetry parameterized as ${\bf \vv}$. The extra Riemann surface $\Tbm$ has one minimal puncture with the parameter $\Al$ and two maximal punctures of the color $n=0$ with the parameters ${\bf \uu}$ and ${\bf \vv}$ respectively. $(b)$: We connect the maximal punctures in $\Tcal_{2,N,\ell}$ and $\Tbm$. It corresponds to gauging the diagonal combination of the $SU(N)^2$ flavor symmetries associated to the two punctures.}
\label{fig:glue}
\end{figure}
In order to capture the surface defects, the authors in \cite{Gaiotto:2015usa} introduced the following procedure. First they prepared the Riemann surface with two maximal punctures of the same color and one minimal puncture, which is denoted by $\Tbm$, \footnote{
Here, by a slight abuse of the notation,
$\Tcal$ means the Riemann surface that gives rise to the 4d theory denoted by the same notation. The subscript of $\Tbm$ comes from the choice for the pole we will make 
in (\ref{eq:Tbm})} in Figure \ref{fig:glue}$(a)$. Next they glued it to $\Tcal_{2,N,\ell}$ through one of the maximal punctures of $\Tcal_{2,N,\ell}$ as described in Figure \ref{fig:glue}$(b)$. Then they closed the minimal puncture labelled by $\Al$. Closing this puncture corresponds to removing all the multiplets with non-zero $U(1)_{\Al}$ charges. The resulting Riemann surface is the same as $\Tcal_{2,N,\ell}$, including the color of the maximal punctures, but one can introduce surface defects labelled by positive inetegers $\rr$, depending on how to close the minimal puncture.
For simplicity, let us focus on the case where $\Be=\Ga=1$ for the moment. If we set $\Al={\T}^{\f{1}{2}}{\q}^{\f{\rr}{N}} (\rr \geq 0)$, the contribution from the chiral multiplet with the charge $\Al^{-1} {\T}^{\f{1}{2}}$ to the index becomes divergent. In the process of removing this chiral multiplet, if we take the residue at the pole $\Al={\T}^{\f{1}{2}}$, we will obtain the index for the original theory $\Tcal_{2,N,\ell}$ without a surface defect at the end. If we take the residue at the pole $\Al={\T}^{\f{1}{2}}{\q}^{\f{\rr}{N}} \ (\rr\geq 1)$, we will obtain the index for the same theory with the surface defect labelled by the $\rr$-th symmetric representation. 

This prescription has the following physical interpretation based on the RG-flow.
First in the UV they added extra chiral multiplets corresponding to $\Tcal^{(++)}_{\Be,-}$.
Let us denote the baryonic operator made from the one with $\Al^{-1} {\T}^{\f{1}{2}}$  as $B$. 
Taking the residue at the former pole amounts to giving it a constant vev. In the IR below the energy scale set by the vev, we flow to the original theory.
Choosing the latter pole corresponds to giving it a position-dependent vev of the form $B(z)=z^r$, where $z$ is a complex coordinate on a plane transverse to the surface defect, since the pole we chose is where the contribution from the operator $(\partial_z)^r B$ is divergent. 
The degree $r$ corresponds to a vortex number.
In the IR, we obtain the original theory in the presence of the surface defect, which is an infinite tension limit of the vortex.

In order to implement this procedure, we have to obtain the index for $\Tbm$ by closing the minimal puncture corresponding to $\De$ in $\Tpp$. First we remove the chiral multiplet with charges $\T^{\f{1}{2}}\De^{-1} \Be^{-1}$ by taking the residue at $\De=\T^{\f{1}{2}} \Be^{-1}$ in the index $\Ical_{\Tpp}$.\footnote{Alternatively we can choose the pole $\De={\T}^{\f{1}{2}} \Be, \, \De={\T}^{-\f{1}{2}}\Ga$ or $\De={\T}^{-\f{1}{2}} \Ga^{-1}$. Each choice gives rise to different surface defects as given in \cite{Gaiotto:2015usa}. In this paper we set $\Be=\Ga=1$ later, so this choice does not matter.} 
Since it corresponds to giving a vev to this chiral multiplet, the cubic superpotentials involving this multiplet become mass terms for the other chiral multiplets and they can be integrated out. Thus the multiplet with charges $\T^{\f{1}{2}}\De^{-1} \Be$ is the only remaining one among the ones with non-zero $U(1)_{\delta}$ charges.
 Let us denote the baryonic operator made from this multiplet by $B$. In order to remove it, we add a chiral multiplet $b$ and couple it to $B$ by a superpotential $bB$. It amounts to calculating the following index 
\begin{align}
\Ical_{\Tbm}:= \Ge (\p \q \Be^{-2N})\, N \Ical_V \, {\rm Res}_{\De={\T}^{\f{1}{2}} \Be^{-1} } \, \f{1}{\De}\, 
\Ical_{  \Tpp } \, ,
\label{eq:Tbm}
\end{align}
where $\Ge (\p \q \Be^{-2N})$ is the contribution from the multiplet $b$.
When we perform the $\zz^{(1)}_{\an}$ contour integral in (\ref{eq:indT2N2}), non-zero contributions to (\ref{eq:Tbm}) come from the following poles 
\begin{align}
\zz^{(1)}_{\an}=\vv^{(2)}_{\sigma({\an})}\, \f{{\T}^{\f{1}{2}}}{\De\Be} \, , \ \ \ ({\an}=1,\cdots, \, N-1) 
\label{eq:zpole}
\end{align}
where $\sigma$ is an element of the symmetric group $S_N$. 
The value of $\zz^{(1)}_N$ is determined from the constraint $\prod_{\an=1}^N \zz^{(1)}_{\an}=1$. Substituting the value to $\Ge({\T}^{\f{1}{2}} v^{(2)}_{\sigma(N)} (z^{(1)}_N)^{-1} \De^{-1} \Be^{-1})$, we see that this factor has a pole $\De={\T}^{\f{1}{2}} \Be^{-1}$.
Evaluating the residue at the pole and summing over the contributions from the poles in $\zz^{(1)}$ classified by permutations $\sigma$, we obtain the following expression
\begin{align}
&\Ical_{\Tbm}=
\Ge (\p \q \Be^{-2N}) 
\,  \nn \\
&\hspace{1em}  \f{\Ical_V^{N-1}}{ N!}  
\oint \prod_{{\an}=1}^{N-1} \f{d\zz^{(2)}_{\an}}{2\pi i \zz^{(2)}_{\an}} 
\f{
\prod_{{\an},{\bn}}
\Ge( \f{\p\q}{{\T} \Be\Ga} (\vv^{(2)}_{{\an}})^{-1} \zz^{(2)}_{\bn}) 
 }{
 \prod_{{\an} \neq {\bn}}
 \Ge(\zz^{(2)}_{\an} /\zz^{(2)}_{\bn})  
}  
\prod_{{\an},{\bn}}
 \Ge ({\T} \Ga \Be^{-1} \vv^{(2)}_{{\an}} /\vv^{(1)}_{\bn} ) 
\, \Ge(\Be^2 \vv^{(1)}_{{\an}} /\zz^{(2)}_{\bn}) 
\nn  \\
& \hspace{3em}
\Ge (\T^{\f{1}{2}} \Be \Al^{-1} (\uu^{(1)}_{\an})^{-1} \vv^{(2)}_{{\bn}} ) 
\Ge (\T^{\f{1}{2}} \Ga^{-1} \Al \uu^{(2)}_{\an} /\vv^{(2)}_{{\bn}})  
  \Ge( \f{\T^{\f{1}{2}}}{\Be\Al} (\uu^{(2)}_{\an})^{-1} \zz^{(2)}_{{\bn}} )    
  \Ge(\T^{\f{1}{2}} \Ga \Al\, \uu^{(1)}_{\an} /\zz^{(2)}_{\bn} ) 
 \,  .
\end{align}

Next we glue it to the theories $\Tcal_{2,N,\ell}$, by gauging the diagonal combinations $SU(N)_{\ii}$ of the $SU(N)_{\ii}^{(0)}$ flavor symmetry in $\Tcal_{2,N,\ell}$ and the $SU(N)$ flavor symmetry labelled by $u^{(1-\ii)}$ in $\Tbm$. We also add bifundamental chiral multiplets $\Phi$ of the gauge groups $SU(N)_1$ and $SU(N)_2$.
The resulting index is given as
\begin{align}
&\Ical[\Tcal_{2,N,\ell}+\Tbm]=
\left(\f{ \Ical_V^{N-1}}{N!} \right)^2 \oint  \prod_{{\an},{\bn}=1}^{N-1} \frac{d\uu^{(1)}_{\an}}{2\pi i \uu^{(1)}_{\an}} \frac{d\uu^{(2)}_{\bn}}{2\pi i \uu^{(2)}_{\bn}}
 \nn\\
 &\hspace{2em}
 \times
  \frac{
  \prod_{{\an},{\bn}}
  \Ge(\frac{\p \q}{\T} (\frac{\Be}{\Ga}) \uu^{(2)}_{{\an}} /\uu^{(1)}_{\bn} )
  \, 
  \Ge( \frac{\p \q}{\T} (\frac{\Ga}{\Be}) \uu^{(1)}_{{\an}} /\uu^{(2)}_{\bn}  )
  }{
  \prod_{{\an}\neq {\bn}}
  \Ge( \uu^{(1)}_{\an}  /{\uu^{(1)}_{\bn}}   )  \Ge( \uu^{(2)}_{\an} / \uu^{(2)}_{\bn}   )  
    }    
    \, \Ical_{\Tcal_{2,N,\ell}}({\bf \uu}^{\dagger}; \Be,\Ga)\ \Ical_{\Tbm}({\bf \uu},\Al, {\bf \vv};\Be,\Ga )\, ,
    \label{eq:indglue}
\end{align}
where ${\bf \uu}^{\dagger}:=(\uu^{(2)}_a, \uu^{(1)}_a)$.

Then we close the minimal puncture $\Al$ by taking the residue at $\Al=\T^{\f{1}{2}} \Be {\q}^{\f{\rr}{N}}$, 
in the above expression. Thus the index in the presence of the surface defect labelled by $\rr$ is given by
 \begin{align}
&\Ical[\Tcal_{2,N,\ell}, \mathfrak{S}_{\rr}]=N \Ical_V\, 
\Ge(\p \q \Be^{2N})
\, 
{\rm Res}_{\Al=\T^{\f{1}{2}}\, \Be \, \q^{\f{\rr}{N}}} \, \f{1}{\Al} \Ical[\Tcal_{2,N,\ell}+\Tbm] \, .
\label{eq:surfaceRes}
\end{align}
When we perform the $\uu^{(1)}_{\an}$ contour integral in (\ref{eq:indglue}), non-zero contributions to (\ref{eq:surfaceRes}) come from the following poles 
\begin{align}
\uu^{(1)}_{\an}= \vv^{(2)}_{ \sigma^{\prime} ( \an)} \f{{\T}^{\f{1}{2}}\Be}{\Al} {\q}^{\rr^{(2)}_{\an}}  \, , \ \ (\an=1,\cdots,N-1)
\end{align}
where $\sigma^{\prime}\in S_N$, $\rr^{(2)}_{\an}\geq 0$ and $\rr^{(2)}_N:=\rr-\sum_{\an=1}^{N-1} \rr^{(2)}_{\an} \geq 0$.
The value of $\uu^{(1)}_N$ is determined from the constraint $\prod_{\an=1}^N \uu^{(1)}_{\an}=1$. Substituting the value to $\Ge({\T}^{\f{1}{2}} \Be \Al^{-1}  (\uu^{(1)}_{N})^{-1} v^{(2)}_{\sigma^{\prime}(N)} )$, we see that this factor has a pole $\Al={\T}^{\f{1}{2}} \Be {\q}^{\f{\rr}{N}}$.

We also have to perform the contour integrals in $\zz^{(2)}$ and $\uu^{(2)}$. As in the case of $\zz^{(1)}$ and $\uu^{(1)}$, 
poles with non-zero residues are classified by elements of $S_N$. For example, the following pole is labelled by the trivial element of $S_N$ both for $\zz^{(2)}$ and $\uu^{(2)}$
\begin{align}
&\zz^{(2)}_{\an}=\Be^2 \vv^{(1)}_{\an} \, , \ \ \ \uu^{(2)}_{\an}=  \f{{\T}^{\f{1}{2}}}{\Be\Al} \zz^{(2)}_{\an} {\q}^{\rr^{(1)}_{\an}} \, , \ (\an=1,\cdots,N-1) \nn
\end{align}  
where $\rr^{(1)}_{\an} \geq 0$ and  $\rr^{(1)}_N:=\rr-\sum_{\an=1}^{N-1} \rr^{(1)}_{\an} \geq 0.$\footnote{If we choose a pole such that $\rr^{(1)}_N<0$, $\Ge({\p}{\q} \Be^{2N}) \Ge({\T}^{\f{1}{2}}\Be^{-1}\Al^{-1} (\uu^{(2)}_N)^{-1} \zz^{(2)}_N )$ becomes zero at the pole when we set $\Be=1$. Also if we choose a pole $\zz^{(2)}_a=\Be^2 \vv^{(1)}_a {\q}^{k_a}$ where $k_a>0$ for some $a\in \{1,\cdots, N-1\}$, $\Ge({\p}{\q} \Be^{-2N}) \Ge(\Be^2 \vv^{(1)}_N / \zz^{(2)}_N )$ is equal to zero at the pole after setting $\Be=1$.} Summing over the residues at all the poles, we can evaluate the index.

From now on, we restrict to the case where $\Be=\Ga=1$. Using the relations (\ref{eGinv}) and (\ref{GamThe}) and the fact that $\Ical_{2,N,\ell}(\uu^{(1)}_{\an}, \uu^{(2)}_{\an})$ $=\Ical_{2,N,\ell}(\uu^{(1)}_{\sigma(\an)}, \uu^{(2)}_{\sigma^{\prime}(\an)})$ for any elements $\sigma$, $\sigma^{\prime}\in S_N$, we obtain the following expression
\begin{align}
&\Ical[\Tcal_{2,N,\ell}, \mathfrak{S}_{\rr}]=
\sum_{ \sum_{\an}\rr^{(1)}_\an=\rr } \sum_{ \sum_{\an}\rr^{(2)}_\an=\rr }\prod_{\ii=1}^2 \prod_{\an,\bn=1}^N
  \f{
\prod_{\lr^{(\ii)}_{\an}=1}^{\rr^{(\ii)}_{\an}} \theta( {\T} (\vv^{(\ii)}_\an)^{-1} \vv^{(\ii+1)}_{\bn}  {\q}^{\rr^{(\ii+1)}_{\bn}-\lr^{(\ii)}_{\an}} ;{\p} )
 }{
     \prod_{\lr^{(i)}_{\bn}=0}^{ \rr^{(i)}_{\bn}-1 } \theta( (\vv^{(i)}_{\an})^{-1} {\vv^{(i)}_{\bn}}  {\q}^{\lr^{(i)}_{\bn}-\rr^{(i)}_{\an}}  ; {\p} )
 }\nn\\
 &\hspace{6em} \times \Ical_{\Tcal_{2,N,\ell}} (\vv^{(1)}_{\an} {\q}^{\rr^{(1)}_\an-\f{\rr}{N}} , \vv^{(2)}_{\an} {\q}^{\rr^{(2)}_{\an}-\f{\rr}{N}})\, ,
 \label{indres:sur}
\end{align}
where $\ii$ is taken modulo $2$ and the theta function $\theta(x;q)$ is defined as (\ref{defTh}).
\\

\begin{figure}[tb]
\begin{center}
\includegraphics[width=16cm]{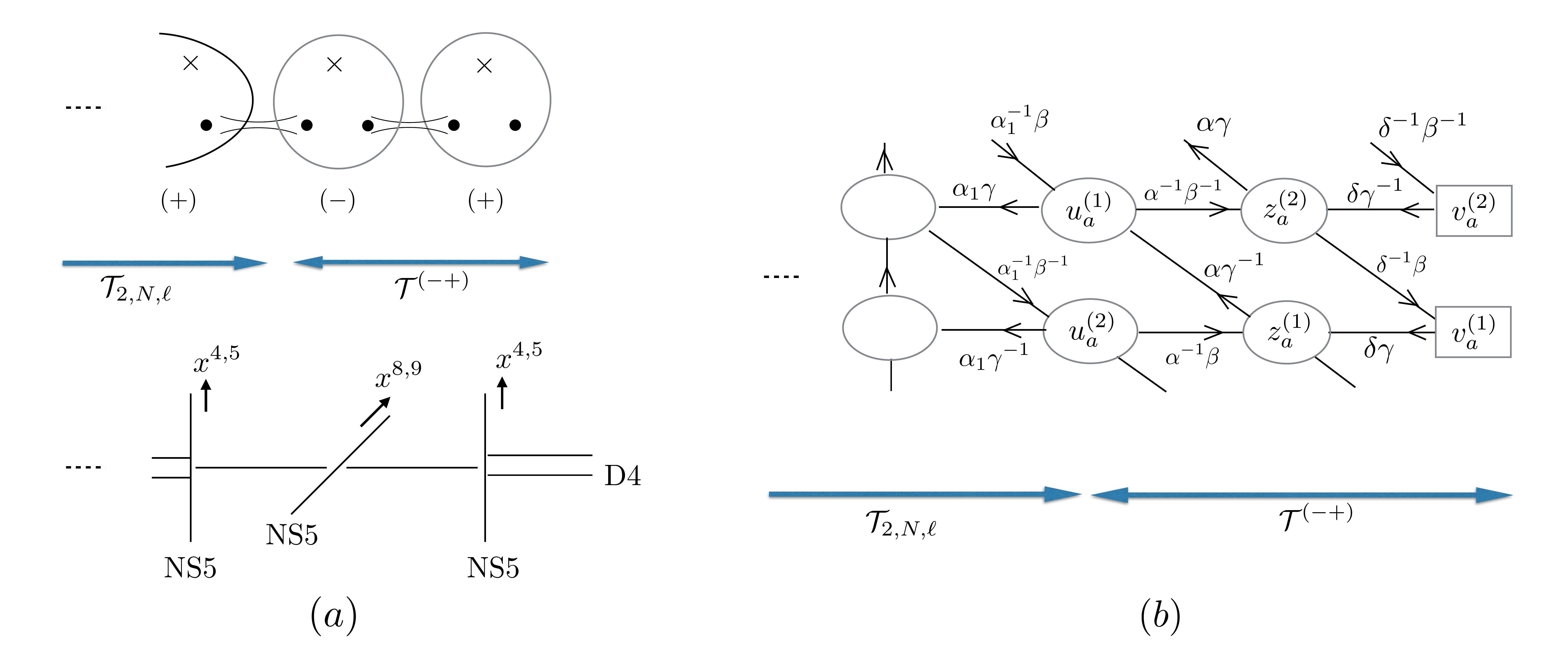}
\end{center}
\vspace{-0.5cm}
\caption{Construction of the second type of surface defects. $(a)$: Riemann surface description for theories after gluing and the corresponding Type I\hspace{-0.1em}IA brane configuration. $(b)$: The quiver diagram for the 4d gauge theories realized by $(a)$.}
\label{fig:quartic}
\end{figure}

Next we insert a different type of surface defects into the same theories $\Tcal_{2,N,\ell}$. In the previous case, we started with the theory $\Tpp$, reduced it to $\Tbm$ and glued it to $\Tcal_{2,N,\ell}$. Instead of $\Tpp$, we can start from $\Tmp$, which is constructed from one trinion with a $-$ sign and the other with a $+$ sign, as described in Figure \ref{fig:quartic}(a). In Type ${\rm I\hspace{-0.1em}IA}$ brane configurations, a minimal puncture in a trinion with a $-$ sign corresponds to an NS5-brane that extends along $012389$. In the 4d gauge theories corresponding to the brane configurations, there are no chiral multiplets $\Phi$, drawn as vertical lines in the quiver diagram, for the gauge groups coming from the D4-branes between two NS5-branes spanning different directions. Note that the $U(1)_{\T}, U(1)_{\Be}$ and $U(1)_{\Ga}$ charges are assigned such that this theory is free from the gauge anomaly and the quartic superpotentials are neutral under the flavor symmetries.

The index for $\Tmp$ is given as 
\begin{align}
&\Ical_{\Tmp}=
\left( \f{\Ical_V^{N-1}}{N!} \right)^2
 \oint
\prod_{ \an , \bn =1}^{N-1}
\f{d
\zz^{(1)}_{\an}
}{2\pi i
\zz^{(1)}_{\an}
}
\f{d
\zz^{(2)}_{\bn}
}{2\pi i
\zz^{(2)}_{\bn}
}
\f{
 1
}{
\prod_{ \an \neq \bn} 
\Ge(\zz^{(1)}_{\an}/ \zz^{(1)}_{\bn})  
\Ge(\zz^{(2)}_{\an}/ \zz^{(2)}_{\bn})
}
\nonumber\\
&
\prod_{\an , \bn} 
\Ge(\T^{\f{1}{2}} \zz^{(1)}_{\an} {\vv^{(1)}_{\bn}}^{-1} \De\Ga ) 
 \Ge(\T^{\f{1}{2}} \vv^{(2)}_{\an}  {\zz^{(1)}_{\bn}}^{-1} \De^{-1}\Be^{-1} ) 
 \Ge(\T^{\f{1}{2}} \zz^{(2)}_{\an} {\vv^{(2)}_{\bn}}^{-1} \De\Ga^{-1} ) 
 \Ge(\T^{\f{1}{2}}  \vv^{(1)}_{\an} {\zz^{(2)}_{\bn}}^{-1} \De^{-1}\Be )  
 \nonumber\\
&
\Ge( \f{  {\p}^{\f{1}{2}}  {\q}^{\f{1}{2}} }{  \T^{\f{1}{2}}}  {\uu^{(1)}_{\an}}  {\zz^{(1)}_{\bn}}^{-1} \Al \Ga^{-1} )  
 \Ge(\f{  {\p}^{\f{1}{2}}  {\q}^{\f{1}{2}} }{\T^{\f{1}{2}}} \Al^{-1} \Be  {\uu^{(2)}_{\an}}^{-1}  \zz^{(1)}_{\bn}) 
 \Ge(\f{ {\p}^{\f{1}{2}}  {\q}^{\f{1}{2}}  \Al \Ga  }{  \T^{\f{1}{2}} } \uu^{(2)}_{\an}  {\zz^{(2)}_{\bn}}^{-1}  )
 \Ge(   \f{  {\p}^{\f{1}{2}}  {\q}^{\f{1}{2}} }{\T^{\f{1}{2}} \Al  \Be  }  {\uu^{(1)}_{\an}}^{-1}   {\zz^{(2)}_{\bn}}    ) 
 \, .
\label{eq:indTmp}
\end{align}
We take the R-charges of the chiral multiplets with non-zero $U(1)_{\Al}$ charges to be $1$ such that the quartic superpotentials have R-charges $2$.
Then we close the minimal puncture $\De$ as
\begin{align}
\Ical_{\Tmpbm}:= \Ge (\p \q \Be^{-2N})\, N \Ical_V \, {\rm Res}_{\De={\T}^{\f{1}{2}} \Be^{-1} } \, \f{1}{\De}\, 
\Ical_{  \Tmp } \, ,
\label{eq:Tmp2}
\end{align}
%
%
glue it to the theories $\Tcal_{2,N,\ell}$ without adding bifundamental chiral multiplets $\Phi$
\begin{align}
&\Ical[\Tcal_{2,N,\ell}+\Tmpbm]=
\left(\f{ \Ical_V^{N-1}}{N!} \right)^2 \oint  \prod_{{\an},{\bn}=1}^{N-1} \frac{d\uu^{(1)}_{\an}}{2\pi i \uu^{(1)}_{\an}} \frac{d\uu^{(2)}_{\bn}}{2\pi i \uu^{(2)}_{\bn}}
 \nn\\
 &\hspace{2em}
 \times
  \frac{
  1
  }{
  \prod_{{\an}\neq {\bn}}
  \Ge( \uu^{(1)}_{\an}  /{\uu^{(1)}_{\bn}}   )  \Ge( \uu^{(2)}_{\an} / \uu^{(2)}_{\bn}   )  
    }    
    \, \Ical_{\Tcal_{2,N,\ell}}({\bf \uu}; \Be,\Ga)\ \Ical_{\Tmpbm}({\bf \uu},\Al, {\bf \vv};\Be,\Ga )\, ,
    \label{eq:indglue2}
\end{align}
and close the minimal puncture $\Al$
 \begin{align}
&\Ical[\Tcal_{2,N,\ell}, \widetilde{\mathfrak{S}}_{\rr}]=N \Ical_V\, 
\Ge(\p \q \Be^{2N})
\, 
{\rm Res}_{\Al=\T^{-\f{1}{2}}\, \Be \, {\p}^{\f{1}{2}}  {\q}^{  \f{1}{2}+ \f{\rr}{N}}} \, \f{1}{\Al} \Ical[\Tcal_{2,N,\ell}+\Tmpbm] \, .
\label{eq:surfaceRes2}
\end{align}
%
Thus the index with the second type of the surface defects is given as
\begin{align}
&\Ical[\Tcal_{2,N,\ell}, \widetilde{\mathfrak{S}}_{\rr}]=
\sum_{ \sum_{\an}\rr^{(1)}_\an=\rr } \sum_{ \sum_{\an}\rr^{(2)}_\an=\rr }\prod_{\ii=1}^2 \prod_{\an,\bn}
  \f{
\prod_{\lr^{(\ii)}_{\an}=1}^{\rr^{(\ii)}_{\an}} \theta( {\T} (\vv^{(\ii)}_\an)^{-1} \vv^{(\ii+1)}_{\bn}  {\q}^{-\lr^{(\ii)}_{\an}} ;{\p} )
 }{
     \prod_{\lr^{(i)}_{\bn}=0}^{ \rr^{(i)}_{\bn}-1 } \theta( (\vv^{(i)}_{\an})^{-1} {\vv^{(i)}_{\bn}}  {\q}^{\lr^{(i)}_{\bn}-\rr^{(i)}_{\an}}  ; {\p} )
 }\nn\\
 &\hspace{6em} \times \Ical_{\Tcal_{2,N,\ell}} (\vv^{(1)}_{\an} {\q}^{\rr^{(1)}_\an-\f{\rr}{N}} , \vv^{(2)}_{\an} {\q}^{\rr^{(2)}_{\an}-\f{\rr}{N}})\, .
 \label{indres:sur2}
\end{align}


\section{2d $\mathcal{N}=(0,2)$ elliptic genus and 4d-2d coupled system for surface defect}
\label{sec:2d4d}
 Four dimensional $\mathcal{N}=2$ superconformal indices with surface defects have been studied in terms of 4d-2d  coupled system \cite{Gadde:2013ftv}.
See also \cite{Chen:2014rca} for 4d-2d coupled system for certain 4d $\mathcal{N}=1$ theories.    
In this picture,  elliptic genera of 2d $\mathcal{N}=(2,2)$ theories supported on surface defect define  the same difference operators
  constructed by RG-flow argument in \cite{Gaiotto:2012xa} up to the fractional fugacity shifts. 
In this section, we  study 4d-2d couple system for superconformal index of class $\mathcal{S}_k$  with  surface defects. 
Since, we are interested in the case $\beta_i=\gamma_i=1$ obtained in the previous section, 
we turn off the corresponding  flavor fugacities in two dimensions.

\begin{figure}[tb]
\begin{center}
\includegraphics[width=7cm]{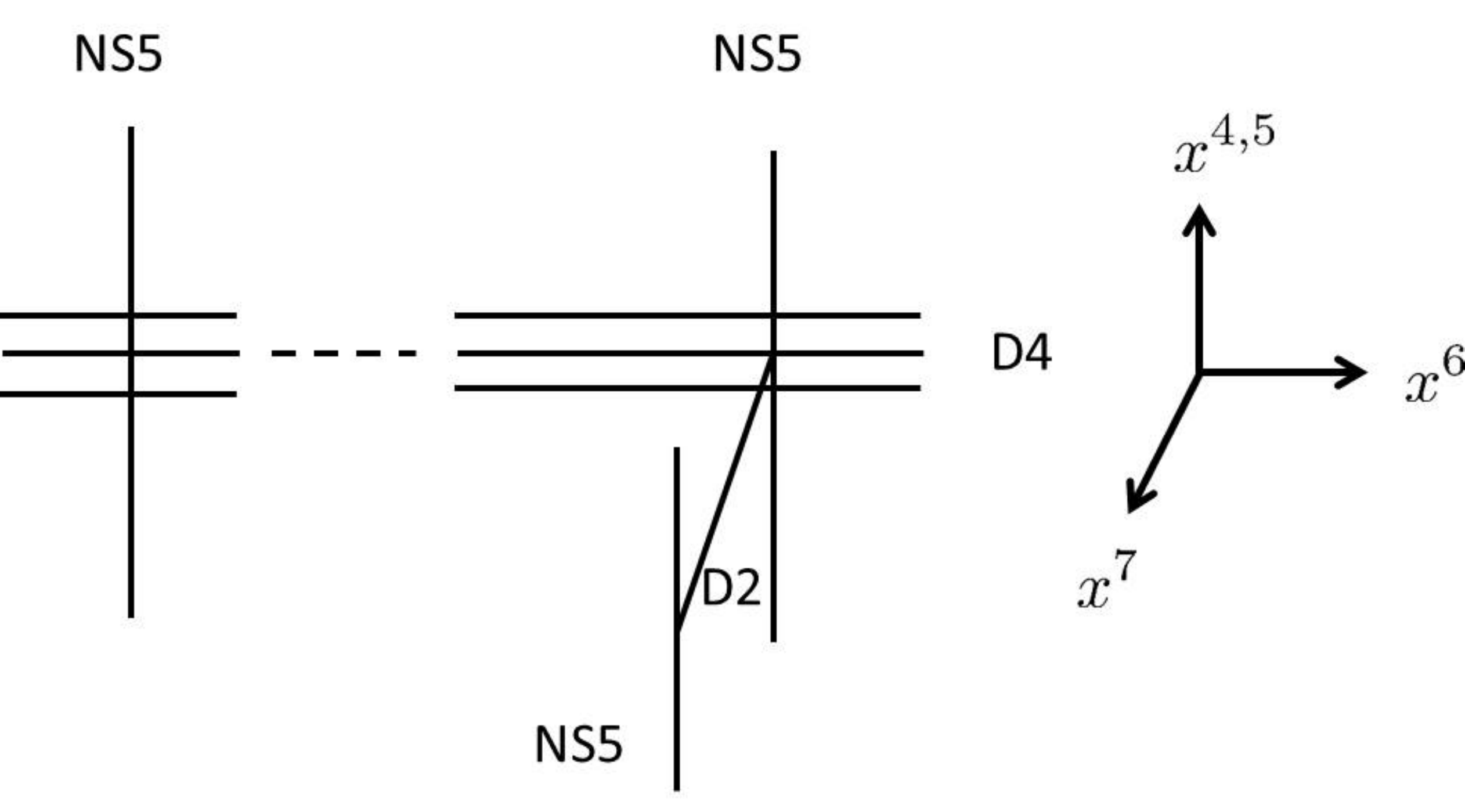}
\end{center}
\vspace{-0.5cm}
\caption{The brane configuration for a half-BPS vortex in 4d $\Ncal=2$ theories.}
\label{fig:surfaceop}
\end{figure}

\begin{table}[htb]
\begin{center}
  \begin{tabular}{|c|c  c c c c c c c| }
  \hline         & ${\mathrm{vec}}^{(i)}$ & $\Sigma^{(i)}$ & $q^{(i)}$    &  $\psi^{(i)}$ & $\tilde{q}^{(i)}$ & $\tilde{\psi}^{(i)}$ &      $\Phi^{(i)}$ & $\Psi^{(i)}$     \\
   \hline $ U(r)_{(i)} \times  U(r)_{(i+1)} $   & $({\rm{ad}}, \mathbf{1})$ & $(\overline{\mathbf{r}}, \mathbf{r})$  
 & $(\mathbf{r},\mathbf{1})$     &  $(\mathbf{1},\mathbf{r})$ & $(\overline{\mathbf{r}},\mathbf{1})$ & $(\mathbf{1},\overline{\mathbf{r}})$  &      
$({\rm{ad}}, \mathbf{1})$ & $(\overline{\mathbf{r}}, \mathbf{r})$ 
  \\
\hline $ SU(N)_{L, (i)} \times  SU(N)_{L, (i+1)} $  & $(\mathbf{1}, \mathbf{1})$ & $(\mathbf{1}, \mathbf{1})$    
 & $(\overline{\mathbf{N}},\mathbf{1})$    &  $(\mathbf{1},\overline{\mathbf{N}})$ & $(\mathbf{1},\mathbf{1})$ & $(\mathbf{1}, \mathbf{1})$  &      
$(\mathbf{1}, \mathbf{1})$ & $(\mathbf{1}, \mathbf{1})$ 
   \\
\hline $ SU(N)_{R, (i)} \times  SU(N)_{R, (i+1)} $ &  $(\mathbf{1}, \mathbf{1})$ & $(\mathbf{1}, \mathbf{1})$     
 & $({\mathbf{1}},\mathbf{1})$    &  $(\mathbf{1},{\mathbf{1}})$  & $({\mathbf{N}},\mathbf{1})$ & $(\mathbf{1},{\mathbf{N}})$   &      
$(\mathbf{1}, \mathbf{1})$ & $(\mathbf{1}, \mathbf{1})$ 
 \\
\hline $ U(1)_c $ & $0$  & $0$   
 & $1$    &  $1$ & $1$ & $1$  &      
$0$ & $0$ 
   \\
\hline $ U(1)_t  $  & $0$ & $1$   
 & $0$    &  $1$ & $0$ & $1$ &      
$0$ & $1$ 
   \\
\hline $ U(1)_d  $  & $0$  & $0$  
 & $0$    &  $0$  & $0$ & $0$  &      
$1$ & $1$ 
  \\  \hline
  \end{tabular}
\end{center}
\caption{The charge assignment for 2d $\mathcal{N}=(0,2)$ theory.}
\label{2dmatter}
\end{table}
Before orbifolding, the
two dimensional  theory associated with a half-BPS surface defect labelled by the $kr$-symmetric representation 
 is $\mathcal{N}=(2,2)$ $U(kr)$ supersymmetric gauge theory  with  an adjoint chiral multiplet, $k N$ fundamental chiral multiplets and $k N$ anti-fundamental chiral multiplets \cite{Gaiotto:2009fs, Gadde:2013ftv}. It is realized by world volume theory of $kr$ D2-branes which describes zero modes of 
half-BPS  vortex  with vortex number $kr$ \cite{Hanany:2003hp}.   
The brane configuration  is specified by Figure \ref{fig:surfaceop}.
$kr$ D2-branes extend along the directions 017 and  suspended by two NS5-branes.  
 Flavor groups $SU(kN) \times SU(kN)$ for the chiral multiplets in two dimensions 
are identified with a flavor and a gauge group in four dimensions associated with $kN$ D4-branes.

Under the orbifold action in \eqref{eq:orbifold} , $\mathcal{N}=(2,2)$ multiplets split into the following $\mathcal{N}=(0,2)$ multiplets.   
The $U(kr)$ $\mathcal{N}=(2,2)$ vector multiplet  split into $\prod_{i=1}^k U(r)_{(i)}$ $\mathcal{N}=(0,2)$ vector multiplets and $k$ bi-fundamental
 chiral multiplets $\Sigma^{(i)}$, $i=1, \cdots, k$ charged under  gauge groups $U(r)_{(i)} \times U(r)_{(i+1)}$ with  $k+1 \equiv 1 ( \mod k )$. $\mathcal{N}=(2,2)$ $U(kr)$ adjoint chiral multiplet split into $k$ $\mathcal{N}=(0,2)$ adjoint chiral multiplets $\Phi^{(i)}$ charged under gauged group $U(r)_{(i)}$  and bi-fundamental chiral multiplets $\Psi^{(i)}$ charged under gauge group $U(r)_{(i)} \times U(r)_{(i+1)}$. The $\mathcal{N}=(2,2)$ fundamental chiral multiplet split into  the fundamental chiral multiplets $q^{(i)}$ and the fundamental Fermi multiplets $\psi^{(i)} , (i=1, \cdots, k )$ . The $\mathcal{N}=(2,2)$ anti-fundamental chiral multiplet split into the anti-fundamental chiral multiplets $\tilde{q}^{(i)}$ and anti-fundamental Fermi multiplets $\psi^{(i)}, (i=1, \cdots, k )$.  
Then 2d $\mathcal{N}=(0,2)$ theory supported on  the half BPS surface defect  is
  $\prod_{i=1}^k U(r)_{(i)}$ circular quiver gauge theory.
 Global symmetry group we concern is $\prod_{i=1}^k SU(N)_{(i), L} \times \prod_{i=1}^k SU(N)_{(i), R} \times U(1)_c \times U(1)_t  \times U(1)_d$. 
 \begin{figure}[tb]
\begin{center}
\includegraphics[width=7cm]{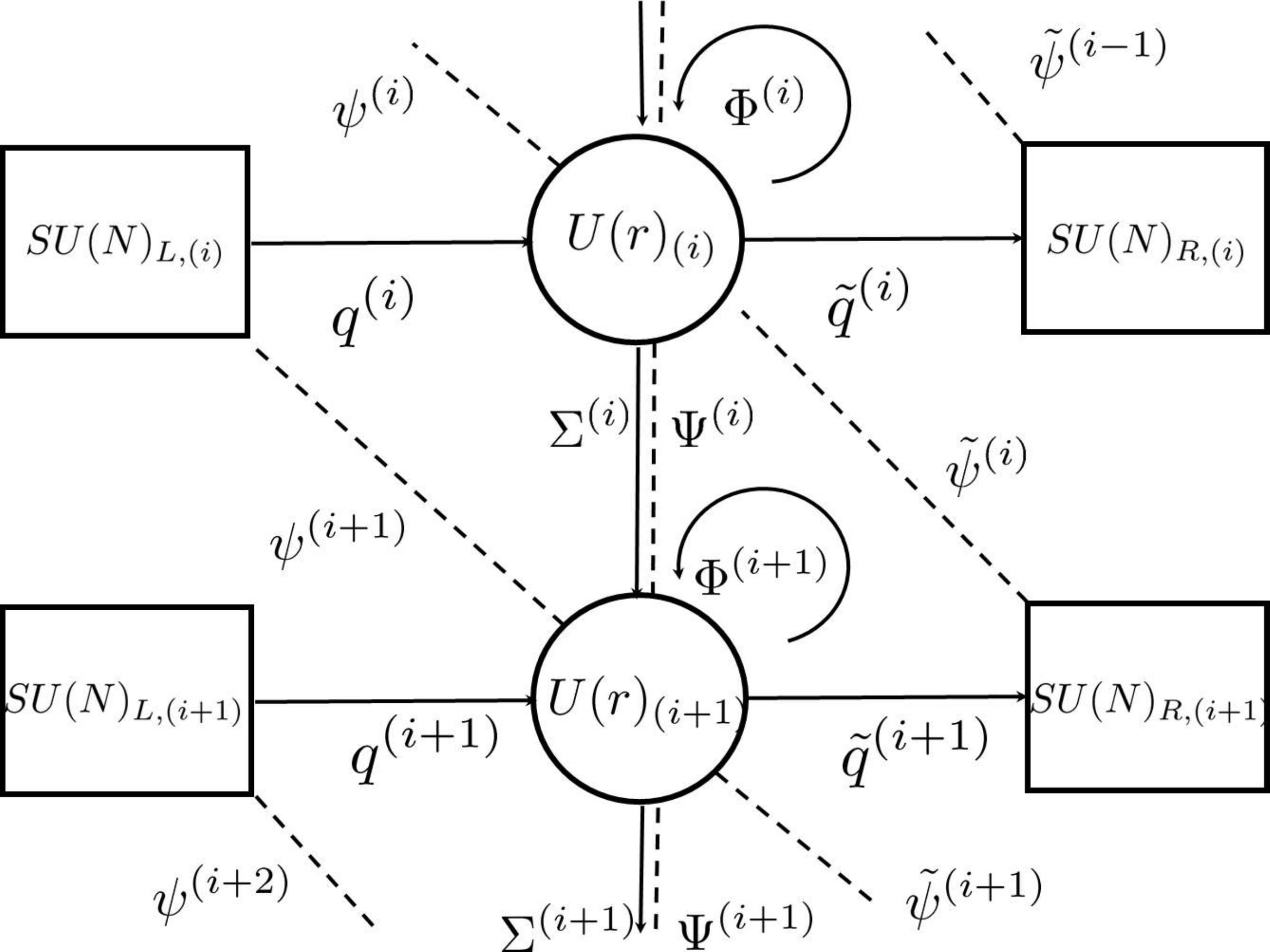}
\end{center}
\vspace{-0.5cm}
\caption{The quiver diagram for the 2d $\mathcal{N}=(0,2)$ theory corresponds to the surface defect ${\mathfrak{S}}_r$. 
The solid lines express the $\mathcal{N}=(0,2)$ chiral multiplets. The dashed lines express $\mathcal{N}=(0,2)$ Fermi multiplets.  }
\label{fig:2d02quiver}
\end{figure} 
 The matter content and quiver are specified in the Table \ref{2dmatter} and Figure  \ref{fig:2d02quiver}.
 $SU(N)_{(i),L}$ in two dimensions  is identified with the  flavor group $SU(N)^{(0)}_i$ of Figure \ref{fig:TkNl} in four dimensions.
 $SU(N)_{(i),R}$ in two dimensions  is identified with the gauge flavor group $SU(N)^{(1)}_{(k-i)}$ of Figure \ref{fig:TkNl} in four dimensions.

We evaluate the elliptic genus for this model.  
The one-loop determinants of $\mathcal{N}=(0,2)$ chiral multiplets are 
\bel
&&Z_{\Sigma}=\prod_{i=1}^k \prod_{\al, \beta=1}^r  \th (t (w^{(i) }_{\al})^{-1} w^{(i+1) }_{\beta} )^{-1}, ~~ 
Z_{\Phi}= \prod_{i=1}^k \prod_{\al, \beta=1}^r  \th (d w^{(i) }_{\al} (w^{(i) }_{\beta})^{-1}  )^{-1} , \non
&&Z_{q}=\prod_{i=1}^k \prod_{\al=1}^r \prod_{a=1}^N \th (c w^{(i)}_{\al} ({\zeta}^{(i) }_{a})^{-1})^{-1}, ~~
Z_{\tilde{q}}=\prod_{i=1}^k \prod_{\al=1}^r \prod_{a=1}^N \th (c (w^{(i) }_{\al})^{-1} \tilde{\zeta}^{(i) }_{a})^{-1} \, .
\label{1loopchiral}
\ee
Here  $Z_{\Sigma}$, $Z_{\Phi}$,  $Z_{q}$  and  $Z_{\tilde{q}}$ express
 the contribution of  bi-fundamental , adjoint, fundamental  and anti-fundamental  chiral multiplets, respectively.  We define $\th(x):=\th(x;q)$.
The  one-loop determinant of vector multiplets and  Fermi multiplets  are
\bel
&&Z_{\mathrm{vec}}= (q ;q)^{2 k r } \prod_{i=1}^k  \prod_{\al \neq \beta }^r \th (w^{(i)}_{\al} (w^{(i)}_{\beta})^{-1} ), ~~
Z_{\Psi}=\prod_{i=1}^k \prod_{\al, \beta=1}^r  \th (t d (w^{(i)  }_{\al} )^{-1} w^{(i+1)}_{\beta}  ) , \non
&&Z_{{\psi}}=\prod_{i=1}^k \prod_{\al=1}^r \prod_{a=1}^N \th (c t w^{(i+1) }_{\al}  ({\zeta}^{(i) }_{a})^{-1}),  
~~Z_{\tilde{\psi}}= \prod_{i=1}^k \prod_{\al=1}^r \prod_{a=1}^N \th (c t (w^{(i) }_{\al})^{-1}  \tilde{\zeta}^{(i+1) }_{a}). 
\label{1loopFermi}
\ee
Here $Z_{\mathrm{vec}}$, $Z_{\Psi}$, $Z_{\Phi}$ and $Z_{\Sigma}$  express
 the contribution of vector multiplet and bi-fundamental, fundamental  and anti-fundamental  Fermi multiplets, respectively. 
$\omega^{(i)}_{\alpha}, (\alpha=1,\cdots, r)$ express a gauge holonomy for the Cartan part of $i$-th  gauge group $U(r)_{(i)}$. 
$\zeta^{(i)}_{a} , (a=1,\cdots, N)$ express fugacities for  the $SU(N)_{(i)L}$  flavor symmetry . 
$\tilde{\zeta}^{(i)}_{a} , (a=1,\cdots, N)$ express fugacities for  the $SU(N)_{(i)R}$ flavor symmetry. 
$c,d,t$ express fugacities for  the $U(1)_c \times U(1)_t  \times U(1)_d$ symmetries.
The elliptic genus \cite{Benini:2013nda, Benini:2013xpa} of this model is written as
\bel
\mathcal{I}_{\text{ell}}(\zeta^{(i)}_a, \tilde{\zeta}^{(i)}_a,c,d,t;q)
= 
\frac{1}{|W|}\sum_{\omega_*} {\rm JK \mathchar `-Res} (\mathbf{Q}(\omega_*), \mathbf{\eta}) Z_{1\mathchar `- {\rm loop}}  \, .
\ee
Here $|W|$ expresses the cardinality of Weyl group
 and $Z_{1\mathchar `- {\rm loop}}$ is the product of all  the one-loop determinant in (\ref{1loopchiral}) and (\ref{1loopFermi}).
 We choose the $\mathbf{\eta} \in \mathbb{R}^{kr}$ as ${\eta}=(-1,\cdots,-1)$. 
 Then the poles which contribute to Jeffrey--Kirwan operation are classified as
\bel
w^{(i)}_{\alpha=(a,r^{(i)}_a)}=c \tilde{\zeta}^{(i)}_{a} d^{l^{(i)}_{a} }, \quad l^{(i)}_{i}=0,1,\cdots,  r^{(i)}_{a}-1
\ee
with $\sum_{a=1}^{N} r^{(i)}_{a}=r$.
Then the elliptic genus is given by 
\bel
\mathcal{I}_{\text{ell}}&&=\sum_{i=1}^k \sum_{\sum_{a=1}^N r^{(i)}_a=r} \prod_{i=1}^k \prod_{a,b=1}^N   
 \frac{\prod_{l^{(i)}_a =0}^{r^{(i)}_a-1} \th (t (\tilde{\zeta}^{(i)}_a)^{-1} \tilde{\zeta}^{(i+1)}_{b} d^{r^{(i+1)}_b-l^{(i)}_a })}
{  \prod_{l^{(i)}_b =0}^{r^{(i)}_b-1} \th (\tilde{\zeta}^{(i)}_a (\tilde{\zeta}^{(i)}_{b})^{-1} d^{r^{(i)}_a-l^{(i)}_b })} \non
&& \times \prod_{i=1}^k  \prod_{a,b =1}^N  \prod_{l^{(i)}_a=0}^{r^{(i)}_a-1}
\frac{\th (c^2 t \tilde{\zeta}^{(i+1) }_{a}  ( {\zeta}^{(i) }_{b} )^{-1} d^{l^{(i+1)}_a}) }{ \th (c^2 \tilde{\zeta}^{(i)}_{a} ( {\zeta}^{(i)}_{b} )^{-1} d^{l^{(i)}_a} )} \, .
\ee

Then  the superconformal index of $\mathcal{T}_{k,N,l}$ with the surface defect 
in the 4d-2d coupled system
 is given by 
\bel
&&\mathcal{I}_{{\rm 4d \mathchar `- 2d }}:=\int \prod_{i=1}^k \prod_{a=1}^{N-1} \frac{d z^{(i)}_a}{2\pi i  z^{(i)}_a} Z_{\mathcal{T}_{k,N,l}} (v^{(i)}_a,  z^{(i)}_a, \beta_i=\gamma_i=1 ; \mathbf{p}, \mathbf{q})  \nonumber \\
&& ~~~~~~~~ \times  \mathcal{I}_{\mathrm{ell}} (\zeta^{(i)}_a=(z^{(k-i)}_a)^{-1}, \tilde{\zeta}^{(i)}_a=(v^{(i)}_a)^{-1}, 
c=\frac{\alpha^{\frac{1}{2}}_1 \mathbf{t}^{\frac{1}{4}}}{\mathbf{q}^{\frac{1}{2}}},d=\mathbf{q}^{-1},t=\frac{\mathbf{pq}}{\mathbf{t}}; q=\mathbf{p}) 
\label{eq:4d2d}
\ee
where $z^{(i)}_a, (a=1\cdots, N-1, i=1, \cdots, k)$ are gauge holonomies for $SU(N)^{(1)}_i, i=1,\cdots,k$ in four dimensions and  
 $ Z_{\mathcal{T}_{k,N,l}}$ is the integrand  of superconformal index of  $\mathcal{T}_{k,N,l}$ without surface defect defined as:
\bel
\mathcal{I}_{\mathcal{T}_{k,N,l}}&&
=:\int \prod_{i=1}^k \prod_{a=1}^{N-1} \frac{d z^{(i)}_a}{2\pi i  z^{(i)}_a} Z_{\mathcal{T}_{k,N,l}} ( v^{(i)}_a, z^{(i)}_a, \beta_i; \mathbf{p}, \mathbf{q})  \nonumber \\
&&= \int \prod_{i=1}^k \prod_{a=1}^{N-1} \frac{d z^{(i)}_a}{2\pi i  z^{(i)}_a}
\Gamma_{e}(\mathbf{t}^{\frac{1}{2}} v^{(i+1)}_{a} (z^{(k-i)}_b)^{-1} \alpha^{-1}_1 \beta_{i+1})  
\Gamma_{e}(\mathbf{t}^{\frac{1}{2}}   (v^{(i) }_{a} )^{-1} {z}^{(k-i) }_{b}  \alpha_1 \gamma^{-1}_{i+1} ) \cdots . \nonumber \\
\ee
Here 
the ellipse denotes the  $v^{(i) }_{a}$ independent terms.
By using the following relations,  
\bel
&&\Gamma_{e}(\mathbf{t}^{\frac{1}{2}}  \alpha^{-1}_1  \mathbf{q}^{r^{(i+1)}_a} v^{(i+1) }_{a}  ({z}^{(k-i) }_{b})^{-1} ) 
 \nonumber \\
&&~~~~~~~~~= \Gamma_{e}(\mathbf{t}^{\frac{1}{2}} \alpha^{-1}_1 v^{(i+1)}_{a} (z^{(k-i)}_b)^{-1} )   \prod_{l^{(i+1)}_a=0}^{r^{(i+1)}_a-1}  
\th (  \mathbf{p} \mathbf{t}^{-\frac{1}{2}} \mathbf{q}^{-l^{(i+1)}_a} (v^{(i+1) }_{a})^{-1}   {z}^{(k-i) }_{b}  \alpha_1)\, ,
  \\ 
&&\Gamma_{e}(\mathbf{t}^{\frac{1}{2}}   (\mathbf{q}^{r^{(i)}_a} v^{(i) }_{a} )^{-1} {z}^{(k-i) }_{b}  \alpha_{1} ) \nonumber \\
&&~~~~~~~~~=  \Gamma_{e}(\mathbf{t}^{\frac{1}{2}}   ( v^{(i) }_{a} )^{-1} {z}^{(k-i) }_{b}  \alpha_1 ) 
\prod_{l^{(i)}_a=1}^{r^{(i)}_a} \th ( \alpha_1 \mathbf{t}^{\frac{1}{2}}  \mathbf{q}^{-l^{(i)}_a}  (v^{(i)}_{a})^{-1} {z}^{(k-i)}_{b}  )^{-1},
\label{ellgammaID}
\ee
we obtain  the following expression of  the  4d-2d coupled index  \eqref{eq:4d2d}  as
\bel
\mathcal{I}_{{\rm 4d \mathchar `- 2d }}=
 \sum_{\sum_{a=1}^N r^{(i)}_a=r} \prod_{i=1}^k \prod_{a,b=1}^N   
 \frac{\prod_{l^{(i)}_a =1}^{r^{(i)}_a} \th (\mathbf{t} ({v}^{(i)}_a)^{-1} v^{(i+1)}_{b} \mathbf{q}^{r^{(i+1)}_b-l^{(i)}_a })}
{  \prod_{l^{(i)}_b =0}^{r^{(i)}_b-1} \th ( (v^{(i)}_a )^{-1} v^{(i)}_{b} \mathbf{q}^{l^{(i)}_b -r^{(i)}_a})} 
\mathcal{I}_{\mathcal{T}_{k,N,l}}  ( v^{(i)}_a \mapsto \mathbf{q}^{r^{(i)}_a} v^{(i)}_a) \, . 
\nonumber \\
\label{eq:2d4dell}
\ee
This index \eqref{eq:2d4dell} for $k=2$ agrees with $\mathcal{I}[\mathcal{T}_{2,N,l}, \mathfrak{S}_r]$ 
\eqref{indres:sur}  up to the overall fractional fugacity shift $\mathbf{q}^{-\frac{r}{N}}$.
When  $\beta_i=\gamma_i=1$, we expect  that the superconformal index for $\mathcal{T}_{k,N,l}$ with surface defect $\mathfrak{S}_r$  is given by
\bel
\mathcal{I}[\mathcal{T}_{k,N,l}, \mathfrak{S}_r]=
\sum_{\sum_{a=1}^N r^{(i)}_a=r} \prod_{i=1}^k \prod_{a,b=1}^N   
 \frac{\prod_{l^{(i)}_a =1}^{r^{(i)}_a} \th (\mathbf{t} ({v}^{(i)}_a)^{-1} v^{(i+1)}_{b} \mathbf{q}^{r^{(i+1)}_b-l^{(i)}_a })}
{  \prod_{l^{(i)}_b =0}^{r^{(i)}_b-1} \th ((v^{(i)}_a )^{-1} v^{(i)}_{b} \mathbf{q}^{l^{(i)}_b -r^{(i)}_a})} 
\mathcal{I}_{\mathcal{T}_{k,N,l}}  ( v^{(i)}_a  \mapsto \mathbf{q}^{r^{(i)}_a -\frac{r}{N}} v^{(i)}_a) \, . \nonumber \\
\label{eq:SCIdefsym}
\ee 
Note that $v^{(i)}_a  \mapsto \mathbf{q}^{r^{(i)}_a -\frac{r}{N}} v^{(i)}_a$ preserves $\prod_{a=1}^N v^{(i)}_a=1$.
If we take $k=1$, \eqref{eq:SCIdefsym} reproduces the difference operator in the $r$-th symmetric representation defined by a surface defect  in  4d $\mathcal{N}=2$ supersymmetric theory.

\begin{figure}[tb]
\begin{center}
\includegraphics[width=7cm]{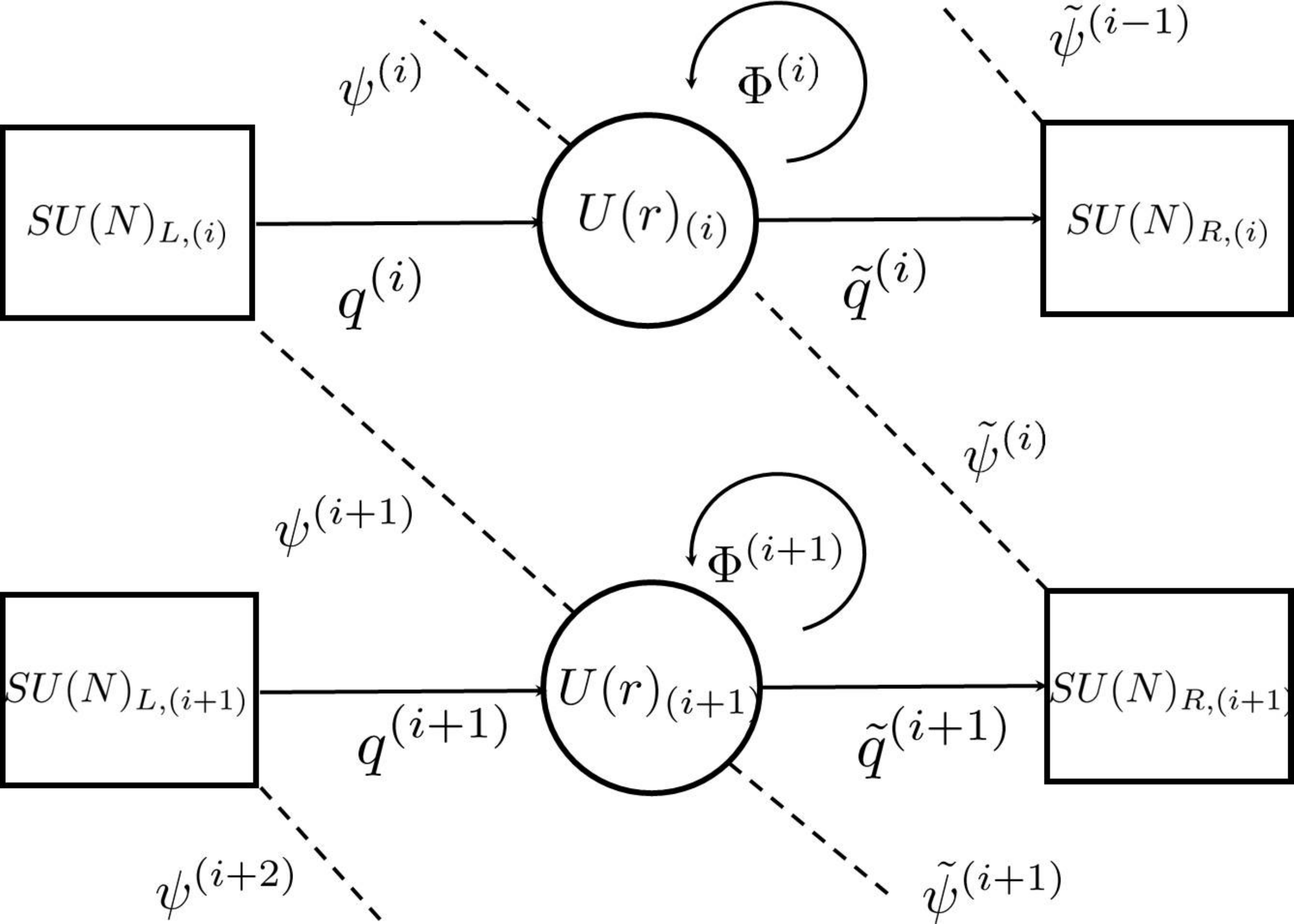}
\end{center}
\vspace{-0.5cm}
\caption{The quiver diagram for the 2d $\mathcal{N}=(0,2)$ theory corresponds to the surface defect $\widetilde{\mathfrak{S}}_r$. The solid lines express the $\mathcal{N}=(0,2)$ chiral multiplets. The dashed lines express $\mathcal{N}=(0,2)$ Fermi multiplets. }
\label{fig:2d02quiver2}
\end{figure}

Next, we study 4d-2d  coupled picture of surface defects in (\ref{indres:sur2}).
The two dimensional theory   is 2d $\mathcal{N}=(0,2)$ $\prod_{i=1}^k U(r)_{(i)}$ quiver gauge theory. The quiver is specified in Figure \ref{fig:2d02quiver2}. This quiver is obtained  by 
removing all the bi-fundamental multiplets from the quiver Figure \ref{fig:2d02quiver}.
Again, the poles which contribute to residue operation are given by
\bel
w^{(i)}_{\alpha=(a,r^{(i)}_a)}=c \tilde{\zeta}^{(i)}_{a} d^{l^{(i)}_{a} }, \quad l^{(i)}_{i}=0,1,\cdots,  r^{(i)}_{a}-1.
\ee
 Then the elliptic genus is written as
\bel
\tilde{\mathcal{I}}_{\text{ell}}&&=\sum_{i=1}^k \sum_{\sum_{a=1}^N r^{(i)}_a=r} \prod_{i=1}^k \prod_{a,b=1}^N   
 \frac{\prod_{l^{(i)}_a =0}^{r^{(i)}_a-1} \th (t (\tilde{\zeta}^{(i)}_a)^{-1} \tilde{\zeta}^{(i+1)}_{b} d^{-l^{(i)}_a })}
{  \prod_{l^{(i)}_b =0}^{r^{(i)}_b-1} \th (\tilde{\zeta}^{(i)}_a (\tilde{\zeta}^{(i)}_{b})^{-1} d^{r^{(i)}_a-l^{(i)}_b })} \non
&& \times \prod_{i=1}^k  \prod_{a,b =1}^N  \prod_{l^{(i)}_a=0}^{r^{(i)}_a-1}
\frac{\th (c^2 t \tilde{\zeta}^{(i+1) }_{a}  ( {\zeta}^{(i) }_{b} )^{-1} d^{l^{(i+1)}_a}) }{ \th (c^2 \tilde{\zeta}^{(i)}_{a} ( {\zeta}^{(i)}_{b} )^{-1} d^{l^{(i)}_a} )} \, .
\ee
The superconformal index in 4d-2d coupled system is given by
\bel
&&\tilde{\mathcal{I}}_{{\rm 4d \mathchar `- 2d }}:=\int \prod_{i=1}^k \prod_{a=1}^{N-1} \frac{d z^{(i)}_a}{2\pi i  z^{(i)}_a} Z_{\mathcal{T}_{k,N,l}} (v^{(i)}_a,  z^{(i)}_a, \beta_i=\gamma_i=1 ; \mathbf{p}, \mathbf{q})  \nonumber \\
&& ~~~~~~~~ \times  \tilde{\mathcal{I}}_{\mathrm{ell}} (\zeta^{(i)}_a=(z^{(k-i)}_a)^{-1}, \tilde{\zeta}^{(i)}_a=(v^{(i)}_a)^{-1}, 
c=\frac{\alpha^{\frac{1}{2}}_1 \mathbf{t}^{\frac{1}{4}}}{\mathbf{q}^{\frac{1}{2}}},d=\mathbf{q}^{-1},t=\frac{\mathbf{pq}}{\mathbf{t}}; q=\mathbf{p}) \, .
\label{eq:4d2d2}
\ee
By using \eqref{ellgammaID},  $\tilde{\mathcal{I}}_{{\rm 4d \mathchar `- 2d }}$
 can be written as
\bel
\tilde{\mathcal{I}}_{{\rm 4d \mathchar `- 2d }}=
 \sum_{\sum_{a=1}^N r^{(i)}_a=r} \prod_{i=1}^k \prod_{a,b=1}^N   
 \frac{\prod_{l^{(i)}_a =1}^{r^{(i)}_a} \th (\mathbf{t} ({v}^{(i)}_a)^{-1} v^{(i+1)}_{b} \mathbf{q}^{-l^{(i)}_a })}
{  \prod_{l^{(i)}_b =0}^{r^{(i)}_b-1} \th ( (v^{(i)}_a )^{-1} v^{(i)}_{b} \mathbf{q}^{l^{(i)}_b -r^{(i)}_a})} 
\mathcal{I}_{\mathcal{T}_{k,N,l}}  ( v^{(i)}_a \mapsto \mathbf{q}^{r^{(i)}_a} v^{(i)}_a) \, .
\label{eq:2d4dell2}
\ee
When $k=2$, this expression agrees with (\ref{indres:sur2}) up to fractional fugacity shift $\mathbf{q}^{-\frac{r}{N}}$.


\section{TQFT structure}
\label{sec:tqft}
In this section, 
we study the 2d TQFT structure of the index for class $\Scal_{k=2}$ theories of type $A_{N-1}$.
The existence of the 2d TQFT structure can be interpreted from the relation between class $\Scal_{k}$ theories on $S^1\times S^3$ and the 2d theories obtained by compactification of the twisted 6d (1,0) theories on $S^1\times S^3$.
First we review that the assumption of the 2d TQFT structure with diagonal structure constants leads to a relation between the eigenfunctions and the eigenvalues of the difference operators $\mathfrak{S}_{\rr}$ that capture the surface defects (\ref{indres:sur}) defined as
\begin{align}
\mathfrak{S}_{\rr} \cdot \Ical_{\Tcal_{2,N,\ell}} (\vt^{(i)}_{\an}) := \Ical [ \Tcal_{2,N,\ell}, \mathfrak{S}_{\rr} ]\, .
\end{align}
Then we will obtain several eigenfunctions and their eigenvalues and check that they satisfy the relation.

Let us start with assuming that the index for class $\Scal_2$ theories can be written in terms of a 2d TQFT correlation function with diagonal structure constants. Namely the index for the trinions with two maximal punctures of the color $n=0$ and $1$ and one minimal puncture can be given as
\begin{align}
&\Ical_{\Tcal_{2, N, 1 }}  (   \ut^{(i)}_{\an}  ,  \Al   , \vt^{(i)}_{\an}   )
=\sum_{\lat} 
 \psi^{[1]}_{\lat}(  {\ut^{(i)}_{\an} }^{-1} ) \, \phi_{\lat}(\Al) \, \psi^{[0]}_{\lat}(\vt^{(i)}_{\an}) \, , \nn\\
 &\Ical_{\Tcal^{(2)}_{2, N, 1 }}  (   \wt^{(i)}_{\an}  ,  \Al   , \ut^{(i)}_{\an}   )
=\sum_{\lat} 
 \psi^{[0]}_{\lat}(  {\wt^{(i)}_{\an} }^{-1} ) \, \phi_{\lat}(\Al) \, \psi^{[1]}_{\lat}(\ut^{(i)}_{\an}) \, ,
\end{align}
where ${\psi}^{[\ell]}_{\lat}$ are assigned to the maximal punctures of the color $\ell$ ($\ell=0,1$). We can write them in terms of the same function $\psi_{\lat}$ as $\psi^{[0]}_{\lat}(  \vt^{(i)}_{\an} ; \Be,\Ga)=\psi_{\lat} (  \vt^{(i)}_{\an} ; \Be,\Ga)$ and $\psi^{[1]}_{\lat}(  \vt^{(i)}_{\an} ; \Be,\Ga)=\psi_{\lat}(  \vt^{(i)}_{\an} ; \Be^{-1},\Ga)$. 
In addition the procedure of gluing these 2d TQFT building blocks is known from the 4d gauge theory side. Gluing the two trinions $\Tcal_{2,N,1}$ and $\Tcal_{2,N,1}^{(2)}$ through the maximal punctures with the parameters $\ut^{(i)}_{\an}$ leads to
\begin{align}
&\Ical_{\Tcal_{2,N,2}}=
\left(  \f{ \Ical_V^{N-1} }{ N!} \right)^2
\oint \prod_{i=1}^2 \prod_{\an=1}^{N-1} \f{d\ut^{(i)}_\an}{2\pi i \ut^{(i)}_\an} \nn\\
& \hspace{4em} \times
\f{ \prod_{\an,\bn}
\Ge( \f{  {\p}{\q} }{ {\T}} {\Be}{\Ga}  \f{ \ut^{(1)}_\an  }{ \ut^{(2)}_\bn }   ) 
\Ge( \f{  {\p}{\q} }{ {\T}}  ({\Be}{\Ga})^{-1}  \f{ \ut^{(2)}_\bn  }{ \ut^{(1)}_\an }   )  
}{
 \prod_{i} \prod_{\an \neq \bn} \Ge(  \ut^{(i)}_\an / \ut^{(i)}_\bn  )
}
\Ical_{ \Tcal^{(2)}_{2,N,1} } (   \wt^{(i)}_{\an}  ,  \Al_2   , \ut^{(1-i)}_{\an}   ) \, 
\Ical_{ \Tcal_{2,N,1} } (   \ut^{(i)}_{\an}  ,  \Al_1   , \vt^{(i)}_{\an}   ) \, . \nn
\end{align}
Thus obtaining the functions $\psi_{\lat}$ and $\phi_{\lat}$ is enough to characterize the 2d TQFT structure.
For simplicity, we focus on the slice $(0,{\q},{\T})$ in the parameter space $({\p},{\q},{\T})$. 
With the help of the expressions for $A_1$ case in \cite{Gaiotto:2015usa}, we expect that the normalized functions $\Pt_{\lat}$ are orthogonal under the following measure $\Dm$ for $A_{N-1}$ case
\begin{align}
& \psi_{\lat}(\vt_{\an}^{(i)}) = K(\vt_{\an}^{(i)}\, ; \Be,\Ga ) \Pt_{\lat} (\vt_{\an}^{(i)} \, ; \Be, \Ga )\, , \ \  
   \nn\\
   & \hspace{2em} K(\vt^{(i)}_\an\, ; 1,1)=\f{1}{
   \prod_{\an, \bn=1}^N  \, 
 ({\T}
  \f{ z^{(1)}_{\an}  }{ z^{(2)}_{\bn} }
  ; {\q} )
   ({\T}
  \f{ z^{(2)}_{\bn}  }{ z^{(1)}_{\an} }
  ; {\q} )
   } \, , 
   \nn\\
&
\left(  \f{ \Ical_V^{N-1} }{ N!} \right)^2
\oint \prod_{i=1}^2 \prod_{\an=1}^{N-1} \f{dz^{(i)}_\an}{2\pi i z^{(i)}_\an} \, \Dm(z^{(i)}_\an)\, \Pt_{\lat}(z^{(i)}_\an \, ; \Be,\Ga) \,  \Pt_{\lat^{\prime}}({z^{(1-i)}_\an}^{-1} \, ; \Be,\Ga) = N_{\lat} \, \delta_{\lat, \, \lat^{\prime}} \, ,  \label{eq:orth} \\
&\hspace{2em} \Dm(z^{(i)}_\an)=
 \f{  \prod_{i=1,2} \prod_{\an \neq \bn}  ( z^{(i)}_{\an}  / z^{(i)}_{\bn}  ; {\q} ) }{
 \prod_{\an, \bn=1}^N  \, 
 ({\T}
 \f{\Be}{\Ga}
  \f{ z^{(1)}_{\an}  }{ z^{(2)}_{\bn} }
  ; {\q} )
   ({\T}
 \f{\Ga}{\Be}
  \f{ z^{(2)}_{\bn}  }{ z^{(1)}_{\an} }
  ; {\q} )
 } \, , \nn
\end{align}
where the q-Pochhammer symbol $(x;q)$ is defined as (\ref{defqP}). 
The prefactor $K$ and the measure $\Dm$ are compatible in the sense that the gluing of the two trinions leads to the following expression
\begin{align}
&\Ical_{\Tcal_{2,N,2}}
%
%
=\sum_{\lat} 
 \psi^{[0]}_{\lat}(  {\wt^{(i)}_{\an} }^{-1} ) \, \phi_{\lat}(\Al_2) N_{\lat} \phi_{\lat}(\Al_1) \, \psi^{[0]}_{\lat}(\vt^{(i)}_{\an}) \, .
\end{align}
Then the index for the 4d theories $\Tcal_{2,N,\ell}$ and $\Tbm$ can be written as
\begin{align}
\label{IR:tqft}
&\Ical_{\Tcal_{2, N, \ell }}  (\vt^{(i)}_{\an}) =
\sum_{\lat} 
 \, \psi^{[\ell]}_{\lat}(  {\wt^{(i)}_{\an} }^{-1} ) \, N_{\lat}^{\ell-1}  \Big(\prod_{m=1}^{\ell} \phi_{\lat}
  (\Al_{m})  \Big) \, \psi^{[0]}_{\lat}(\vt^{(i)}_{\an}) \, , \\
  &\Ical_{\Tcal^{(++)}_{\Be,-}} ( \ut^{(i)}_{\an}, \, \vt^{(i)}_{\an} ) =
\sum_{\lat} C_{\lat}^{(\Be,-)}  \psi^{[0]}_{\lat}({\ut^{(i)}_{\an}}^{-1}) \,  \phi_{\lat} (\Al)  \, \psi^{[0]}_{\lat}(\vt^{(i)}_{\an}) \, . \nn
\end{align}
To engineer the surface defects, we glued the above two theories and the resulting index is given as
\begin{align}
&\Ical[ \Tcal_{2,N,\ell}+ \Tcal^{(++)}_{\Be,-}  ]
=
\sum_{\lat}
\psi^{[\ell]}_{\lat}(  {\wt^{(i)}_{\an} }^{-1} )
  \, \Big(\prod_{m=1}^{\ell} \phi_{\lat} (  \Al_m)  \Big) \,   
N_{\lat}^{\ell} \, C^{(\Be,-)}_{\lat}  \phi_{\lat}(\Al) \, \psi_{\lat} (\vt^{(i)}_{\an}) \, .
\end{align}
Recalling that the difference operators $\mathfrak{S}_{\rr}$ were obtained as in (\ref{eq:surfaceRes}), we see that
\begin{align}
\label{Stqft}
 \mathfrak{S}_{\rr} \cdot  \Ical_{\Tcal_{2,N,\ell}}(\vt^{(i)}_\an)
= N \Ical_V
\sum_{\lat} 
 \psi^{[\ell]}_{\lat}({\wt^{(i)}_\an}^{-1}) \, \Big(\prod_{m=1}^{\ell} \phi_{\lat} (\Al_m) \Big) \,   
N_{\lat}^{\ell} \, C^{(\Be,-)}_{\lat} 
\Big( {\rm Res}_{ {\Al}={\T}^{\f{1}{2}} \Be \, {\q}^{\f{\rr}{N}}  } \phi_{\lat}(\Al) \Big) \psi_{\lat} (\vt^{(i)}_\an) \, .
\end{align}
By using the orthogonality (\ref{eq:orth}), 
we can extract each summand
\begin{align}
\mathfrak{S}_{\rr} \cdot \psi_{\lat} (  \vt^{(i)}_\an )
=  N \Ical_{V} N_{\lat}^{\ell}  C^{(\Be,-)}_{\lat}  \, \Big( {\rm Res}_{ {\Al}={\T}^{\f{1}{2}} \Be \, {\q}^{\f{\rr}{N}}  } \phi_{\lat}(\Al) \Big) \psi_{\lat} (  \vt^{(i)}_\an ) \, .
\end{align}
Using the fact that substituting $\rr=0$ into (\ref{Stqft}) brings us back to (\ref{IR:tqft}), we can rewrite $C^{(\Be,-)}_{\lat}$ in terms of the residue of the function $\phi_{\lat}$. Thus we obtain the following relation
\begin{align}
\mathfrak{S}_{\rr} \cdot \psi_{\lat} (  \vt^{(i)}_\an )
= 
\f{
{\rm Res}_{ {\Al}={\T}^{\f{1}{2}} \Be \, {\q}^{\f{\rr}{N}}  } \phi_{\lat}(\Al) 
}{
{\rm Res}_{ {\Al}={\T}^{\f{1}{2}} \Be  } \phi_{\lat}(\Al) 
}
\,  \psi_{\lat} (  \vt^{(i)}_\an ) \, .
\label{eq:phiRes}
\end{align}

Next we will check that the above relation holds for several eigenfunctions $\psi_{\lat}$. From now on, we restrict to the case $\Be=\Ga=1$. We find that 
the functions $\psi_{\lat}=K P_{\lat}$ with the following normalized parts $P_{\lat}$ are eigenfunctions of $\mathfrak{S}_{\rr=1}$ 
for $SU(N)$ gauge theories
\begin{align}
&\psi_{\lat}(\vt^{(i)}_\an ) =K(\vt^{(i)}_\an \, ;1,1 ) \,  P_{\lat} (\vt^{(i)}_\an \, ; 1,1 ) \, , \nn\\
&\ \ P_{(0)}  (  \vt^{(i)}_\an \, ; 1,1 )   \, = 1  \, , \nn\\
&\ \ P_{(1)_{\pm}}  (\vt^{(i)}_\an \, ; 1,1 )  \,  =
\sum_{\an=1}^N \vt^{(1)}_\an \pm \sum_{\an=1}^N \vt^{(2)}_\an   
 \, ,
\nn\\
&\ \ P_{(2)_{0}}  (\vt^{(i)}_\an \, ; 1,1 )  \,  = 
\Big(   
 \sum_{\an=1}^N (\vt^{(1)}_\an)^2 + \sum_{\an<\bn}  \vt^{(1)}_\an \vt^{(1)}_\bn  
  \Big)
- \Big(  
 \sum_{\an=1}^N  (\vt^{(2)}_\an)^2 + \sum_{\an<\bn}  \vt^{(2)}_\an \vt^{(2)}_\bn
     \Big)  \nn\\
& \hspace{13em}+
\f{{\q}-{\T}^2}{1-{\q}{\T}^2}
\Big(
  \sum_{\an<\bn}  \vt^{(1)}_\an \vt^{(1)}_\bn   -     \sum_{\an<\bn}  \vt^{(2)}_\an b^{(2)}_\bn
\Big) \, .
\label{eq:eigenf}
\end{align}
In case where $N=2$, the eigenvalues $E_{\rr=1 , \lat}$ of $\mathfrak{S}_{\rr=1}$ for each eigenfunctions $\psi_{\lat}$ are as follows
\begin{align}
&E_{1,(0)}=\f{ (1-{\T}^2 )^2 }{  (1-{\q}^{-1})^2 } \, , 
\hspace{9em}
E_{1,(1)_+}=\f{  {\q}^{-\f{1}{2}}  (1-{\T}^2 )  (1-{\T})   ( 1+{\q}{\T} ) 
}{
  (1-{\q}^{-1})^2  
}  
\, , \nn \\
&E_{1,(1)_-}=\f{  {\q}^{-\f{1}{2}}  (1-{\T}^2 )  (1+{\T})   ( 1-{\q}{\T} ) 
}{
  (1-{\q}^{-1})^2  
}  \, ,
\hspace{1em}
E_{1,(2)_0}=
\f{  {\q}^{-1}  (1-{\T}^2 )     ( 1-{\q}^2 {\T}^2 ) 
}{
  (1-{\q}^{-1})^2  
} 
 \, .
\label{eq:ev}
\end{align}
In addition we propose that the functions $\phi_{\lat}(\Al \, ;\Be,\Ga)$ for minimal punctures have the following form
\begin{align}
&\phi_{\lat} (\Al\, ;\Be,\Ga ) = K_{min} ( \Al \, ; \Be,\Ga)  \  {\dd}_{\lat}  \ P^{(min)}_{\lat}( \Al\, ; \Be,\Ga )  \, , \nn \\
&
P^{(min)}_{\lat}
( \Al \, ; 1,1 ) \nn\\
& \ \ 
=
P_{\lat} (
\vt^{(i)}_{\an=1,\cdots,N-1}={\T}^{  N+ \f{3}{2} -i- 2\an } \Al, \, \vt^{(1)}_N={\T}^{-\f{N-1}{2} }  \Al^{-N+1},  \vt^{(2)}_N={\T}^{ \f{N-1 }{2} }  \Al^{-N+1}  ; \Be=\Ga=1  ) \, ,
  \nn \\ 
& K_{min} ( \Al \, ; \Be,\Ga)
=L(\Be,\Ga)  
\prod_{\pm}
\f{1}{
( {\T}^{\f{N}{2}} \Be^{\pm N}  \Al^{-N} ;{\q}  )
( {\T}^{\f{N}{2}} \Ga^{\pm N}  \Al^{N} ;{\q}  )  
} \, , \nn
\end{align}
where ${\dd}_{\lat}$ and $L(\Be,\Ga)$ are $\Al$-independent factors. 
From the above expressions, we can calculate the right-hand side of (\ref{eq:phiRes}) as
\begin{align}
&\Bigg(
\f{
{\rm Res}_{ \Al= {\T}^{\f{1}{2}} \Be \, {\q}^{ \f{\rr}{N} }  }
\, \phi_{\lat} (\Al \, ;\Be,\Ga ) 
}{
{\rm Res}_{ \Al= {\T}^{\f{1}{2}} \Be  }
\, \phi_{\lat} (\Al \, ;\Be,\Ga ) 
}
\Bigg) \Bigg|_{\Be=\Ga=1} 
=\f{ (1-{\T}^N)^2 }{  ( 1-{\q}^{-1} )^2 }  \
\f{
P^{(min)}_{\lat} (  {\T}^{\f{1}{2}}  {\q}^{ \f{\rr}{N} }  \, ;1,1 )
}{
P^{(min)}_{\lat} (  {\T}^{\f{1}{2}}  \, ;1,1 )
} \, .
\label{eq:phiResR}
\end{align}
Comparing (\ref{eq:ev}) with (\ref{eq:phiResR}), we find that the relation (\ref{eq:phiRes}) with $\rr=1$ holds for $N=2$. We also checked it for $N=3$ and $4$.
Moreover the relation (\ref{eq:phiRes}) implies that the difference operators $\mathfrak{S}_{\rr}$ with any positive integer $\rr$ have the same eigenfunctions. Indeed we checked that the functions (\ref{eq:eigenf}) are also eigenfunctions of $\mathfrak{S}_{\rr}$ with $\rr=2$ and $3$ for $N=2$ and $\mathfrak{S}_{\rr}$ with $\rr=2$ for $N=3$.

It would be interesting to check that the difference operators $\widetilde{\mathfrak{S}}_{\rr}$ for the second type of surface defects in (\ref{indres:sur2}) also satisfy a relation similar to (\ref{eq:phiRes}). At least we checked that the functions (\ref{eq:eigenf}) are also eigenfunctions of $\widetilde{\mathfrak{S}}_{\rr}$ with $\rr=1,2$ and $3$ for $N=2$ and $\widetilde{\mathfrak{S}}_{\rr}$ with $\rr=1$ for $N=3$.

\section{Conclusion}
\label{sec:conc}
In this paper, we have studied the superconformal index of class $\mathcal{S}_k$ theories with  half-BPS surface defects.
There are two types of  surface defects that we have considered.  One type arises from gluing a trinion with a $+$ sign
 to punctured Riemann surfaces with the same signs representing class $\mathcal{S}_k$ theories, while the other is obtained by gluing a sphere  with a $-$ sign.
 We calculated the index with these surface defects labelled by generic symmetric representations of $su(N)$ in the formalism based on the Riemann surface description as in \cite{Gaiotto:2015usa}.  
This index defines difference operators acting on the superconformal indices of class $\mathcal{S}_k$ theories, labelled by positive integers $r$ corresponding to the symmetric representations.

We have also studied the above half-BPS surface defects in terms of   4d-2d systems. The two dimensional $\mathcal{N}=(0,2)$ gauge
theories denoted by Figure \ref{fig:2d02quiver}, which correspond to the former type of surface defects, are  interpreted as  an orbifold of two-dimensional $\mathcal{N}=(2,2)$ gauge theories associated with half-BPS surface defects in four-dimensional $\mathcal{N}=2$ theories. On the other hand, the two dimensional $\mathcal{N}=(0,2)$ gauge 
theories which we found as in Figure \ref{fig:2d02quiver2} for the latter type of surface defects do not  originate from an orbifold of the $\mathcal{N}=(2,2)$ theories. 
We evaluated the elliptic genus of these two classes of two dimensional $\mathcal{N}=(0,2)$ gauge 
theories and found that the 4d-2d combined index reproduces the superconformal indiex with surface defects obtained as in \cite{Gaiotto:2015usa}  up to the overall fractional fugacity shift.
It would be interesting to study the origin of this shift in the 4d-2d coupled systems.

Moreover we study the 2d TQFT structure of the class $\Scal_{k=2}$ index where a basis of eigenfunctions of the difference operators $\mathfrak{S}_{\rr}$ diagonalize the superconformal index for a trinion with three punctures.
This property is useful to determine the superconformal index of class $\Scal_{k=2}$ theories. 
Several eigenfunctions in the limit $\Be=\Ga=1$ and ${\p}={\q}=0$ have been constructed in terms of the difference operators $\mathfrak{S}_{\rr}$ in (\ref{indres:sur}) for $\rr=1$ \cite{Gaiotto:2015usa}.
 We have obtained eigenfunctions and their eigenvalues in the limit $\Be=\Ga=1, {\p}=0$ but ${\q}\neq 0$ for $\rr=1$ and found that they satisfy the relation that was derived from the 2d TQFT structure with the diagonal structure constants.
In addition, we checked that these eigenfunctions are also eigenfunctions of higher $\rr$ difference operators $\mathfrak{S}_{\rr}$ and $\widetilde{\mathfrak{S}}_{\rr}$, which we have constructed, and it is also consistent with the 2d TQFT structure.

We briefly comment on future directions of our work.
In class $\Scal$ theories, an interesting property of surface defects is that the composition of the two difference operators labelled by the representations $R_1$ and $R_2$ can be decomposed as a sum of the difference operators labelled by the representations that appear in the direct sum decomposition of $R_1\otimes R_2$. It gives Verlinde formula and its $({\q},{\T})$-deformation \cite{Gaiotto:2012xa, Alday:2013kda, Bullimore:2014nla, Aganagic:2011sg}. 
It is interesting to study an algebra of surface defects in class $\mathcal{S}_k$ theories. 
While we restricted to the surface defects labelled by symmetric representations of $su(N)$,
it would be fascinating to understand how to describe the surface defects corresponding to generic representations in terms of 4d-2d coupled systems.

\section*{Acknowledgments}

Y.I. thanks Heeyeon Kim for valuable discussion.
We  are grateful to Takuya Okuda for reading the manuscript and giving useful comments.

\appendix
\section{Elliptic Gamma function and Theta function} 
\label{sec:app}
The elliptic Gamma function is defined as
\begin{align}
&\Gamma(z;{\p}, {\q}):=\prod_{m,n\geq0 } \f{1-{\p}^{1+m} {\q}^{1+n} z^{-1} }{1-{\p}^m {\q}^n z } 
\label{defGam}
\end{align}
and it satisfies the following relation
\begin{align}
\Gamma(z ;{\p}, {\q} )=\Gamma \left( \f{ {\p} {\q} }{z}  ;{\p}, {\q} \right)^{-1} \, .
\label{eGinv}
\end{align}
The theta  function is defined as
\begin{align}
&(x; q):=\prod_{k\geq 0} (1-x q^k) \, , 
\label{defqP} \\
&\theta(x, q):=(x; q) ( q/x; q) \, . 
\label{defTh}
\end{align}
We see that the following relation holds
\begin{align}
\f{ \Gamma ({\q} z ; {\p},{\q} ) }{ \Gamma(z ; {\p},{\q} )  }=\theta(z,{\p}) \, .
\label{GamThe}
\end{align}

\end{document}